\documentclass[a4paper,11pt]{article}
\pdfoutput=1 

\usepackage{jheppub} 
\makeatletter
\DeclareRobustCommand*{\bfseries}{%
  \not@math@alphabet\bfseries\mathbf
  \fontseries\bfdefault\selectfont
  \boldmath
}
\makeatother

\usepackage{calc}
\usepackage{rotating}
\usepackage[english]{babel}
\usepackage{graphicx}
\usepackage{subfig}
\usepackage{float}
\usepackage{amsmath}
\usepackage{amssymb}
\usepackage{amsthm}
\usepackage{latexsym}
\usepackage{dcolumn}
\newcommand{\newc}{\newcommand*}

\long\def\begincomment#1\endcomment{%
        \begingroup\sf\baselineskip12pt#1\endgroup}

\newc{\etal}{\textrm{et al.}} 
\newc{\eg}{\textrm{e.g.}} 
\newc{\ie}{\textrm{i.e.}}
\newc{\etc}{\textrm{etc}}
\newc\vs{\textrm{vs.}}
\newc{\cl}{\rm {C.L.}}

\newc{\ev}{\ensuremath{\,\mathrm{eV}}}
\newc{\kev}{\ensuremath{\,\mathrm{keV}}}
\newc{\mev}{\ensuremath{\,\mathrm{MeV}}}
\newc{\gev}{\ensuremath{\,\mathrm{GeV}}}
\newc{\tev}{\ensuremath{\,\mathrm{TeV}}}
\newc{\MeV}{\mev} 
\newc{\TeV}{\tev}
\newc{\invpb}{\ensuremath{/\text{pb}}}
\newc{\invfb}{\ensuremath{\,\text{fb}^{-1}}}
\newc\nb{\ensuremath{\,\mathrm{nb}}} \newc\pb{\ensuremath{\,\mathrm{pb}}} \newc\fb{\ensuremath{\,\mathrm{fb}}}
\newc\pc{\ensuremath{\,\mathrm{pc}}}
\newc\kpc{\ensuremath{\,\mathrm{kpc}}}
\newc\mpc{\ensuremath{\,\mathrm{Mpc}}}
\newc\ps{\ensuremath{\,\mathrm{ps}}} 
\newc\cmeter{\ensuremath{\,\mathrm{cm}}} 
\newc\meter{\ensuremath{\,\mathrm{m}}} 
\newc\kmeter{\ensuremath{\,\mathrm{km}}}
\newc\second{\ensuremath{\,\mathrm{s}}}
\newc\msecond{\ensuremath{\,\mathrm{ms}}}
\newc\nsecond{\ensuremath{\,\mathrm{ns}}}
\newc\psecond{\ensuremath{\,\mathrm{ps}}}

\newc{\chisqmin}{\ensuremath{\chi^2_{\mathrm{min}}}}
\newc{\Delchisq}{\ensuremath{\Delta\chi^2}}
\newc{\chisq}{\ensuremath{\chi^2}}
\newc{\like}{\ensuremath{\mathcal{L}}}

\newc\lsim{\ensuremath{\mathrel{\rlap{\lower4pt\hbox{\hskip1pt$\sim$}}\raise1pt\hbox{$<$}}}}
\newc\gsim{\ensuremath{\mathrel{\rlap{\lower4pt\hbox{\hskip1pt$\sim$}}\raise1pt\hbox{$>$}}}}
\newc{\VEV}[1]{\ensuremath{\langle #1 \rangle}}
\newc{\dl}{\ensuremath{\stackrel{\leftarrow}{D}}}
\newc{\dr}{\ensuremath{\stackrel{\rightarrow}{D}}}

\newc{\bcenter}{\begin{center}}    \newc{\ecenter}{\end{center}}
\newc{\bfl}{\begin{flushleft}}    \newc{\efl}{\end{flushleft}}
\newc{\bfr}{\begin{flushright}}    \newc{\efr}{\end{flushright}}

\newc{\bi}{\begin{itemize}}
\newc{\ei}{\end{itemize}}
\newc{\bed}{\begin{description}}
\newc{\eed}{\end{description}}
\newc{\ben}{\begin{enumerate}}
\newc{\een}{\end{enumerate}}

\newc{\be}{\begin{equation}}
\newc{\ee}{\end{equation}}
\newc{\bea}{\begin{eqnarray}}
\newc{\eea}{\end{eqnarray}}
\newc{\ra}{\rightarrow}

\newc{\alphas}{\ensuremath{\alpha_s}}
\newc{\alphatwo}{\ensuremath{\alpha_2}}
\newc{\alphaone}{\ensuremath{\alpha_1}}
\newc{\alphai}[1]{\ensuremath{\alpha_{#1}}}
\newc{\alphaem}{\ensuremath{\alpha_{\mathrm{em}}}}
\newc{\alphaeff}{\ensuremath{\alpha_{\mathrm{eff}}}}
\newc{\sineff}{\ensuremath{\sin \theta_{\mathrm{eff}}}}
\newc{\sinsqeff}{\ensuremath{\sin^2 \theta_{\mathrm{eff}}}}
\newc{\dalphahad}{\ensuremath{\Delta \alpha_{\mathrm{had}}}}
\newc{\yt}{\ensuremath{h_t}} \newc{\yb}{\ensuremath{h_b}} \newc{\ytau}{\ensuremath{h_{\tau}}}
\newc\mz{\ensuremath{M_Z}} 
\newc\mw{\ensuremath{m_W}}
\newc\mZ{\mz}        \newc\mW{\mw}
\newc\mhsm{\ensuremath{ m_{H_{\mathrm{SM}}}}}
\newc{\mtop}{\ensuremath{ m_t}}               \newc{\mtpole}{\ensuremath{ M_t}}
\newc{\mbottom}{\ensuremath{ m_b}} 
\newc{\mtau}{\ensuremath{ m_{\tau}}}
\newc{\mt}{\mtpole}
\newc{\mb}{\mbottom} 
\newc{\rgg}{\ensuremath{R_{h}(\gamma\gamma)}}
\newc{\rzz}{\ensuremath{R_{h}(ZZ)}}
\newc{\rtwogg}{\ensuremath{R_{h_2}(\gamma\gamma)}}
\newc{\rtwozz}{\ensuremath{R_{h_2}(ZZ)}}
\newc{\ronegg}{\ensuremath{R_{h_1}(\gamma\gamma)}}
\newc{\ronezz}{\ensuremath{R_{h_1}(ZZ)}}
\newc{\rsiggg}{\ensuremath{R_{h_\textrm{sig}}(\gamma\gamma)}}
\newc{\rsigzz}{\ensuremath{R_{h_\textrm{sig}}(ZZ)}}
\newc{\llbar}{\ensuremath{\ell\bar{\ell}}}
\newc{\tauptaum}{\ensuremath{ \tau^+\tau^-}}
\newc{\qqbar}{\ensuremath{ q\bar{q}}} \newc{\ppbar}{\ensuremath{ p\bar{p}}}
\newc{\bbbar}{\ensuremath{ b\bar{b}}} \newc{\ttbar}{\ensuremath{ t\bar{t}}}
\newc{\ffbar}{\ensuremath{ f\bar{f}}} \newc{\tautaubar}{\ensuremath{ \tau\bar{\tau}}}
\newc{\mchi}{\ensuremath{m_{\chi}}}
\newc{\squark}{\ensuremath{\tilde{q}}}
\newc{\slepton}{\ensuremath{\tilde{l}}}
\newc{\gluino}{\ensuremath{\tilde{g}}} 
\newc{\mgluino}{\ensuremath{{m_{\gluino}}}}
\newc{\tone}{\ensuremath{{\tilde{t}_1}}}

\newc{\sthw}{\ensuremath{ \sin\theta_W}}              \newc{\cthw}{\ensuremath{\cos\theta_W}}
\newc{\tanthw}{\ensuremath{ \tan\theta_W}}              \newc{\cotthw}{\ensuremath{\cot\theta_W}}
\newc{\ssqthw}{\ensuremath{\sin^2 \theta_W}}
\newc{\msbar}{\ensuremath{\overline{MS}}} \newc{\drbar}{\ensuremath{\overline{DR}}}
\newc{\mtmtsmmsbar}{\ensuremath{ m_t(m_t)^{\msbar}_{{\mathrm{SM}}}}}
\newc{\mtmtsmdrbar}{\ensuremath{ m_t(m_t)^{\drbar}_{{\mathrm{SM}}}}}
\newc{\mtmtmssmdrbar}{\ensuremath{ m_t(m_t)^{\drbar}_{{\mathrm{SUSY}}}}}
\newc{\mbmbmsbar}{\ensuremath{ m_b(m_b)^{\msbar} }}
\newc{\mbmbsmmsbar}{\ensuremath{ m_b(m_b)^{\msbar}_{{\mathrm{SM}}}}}
\newc{\mbmzsmmsbar}{\ensuremath{ m_b(\mz)^{\msbar}_{{\mathrm{SM}}}}}
\newc{\mbmzsmdrbar}{\ensuremath{ m_b(\mz)^{\drbar}_{{\mathrm{SM}}}}}
\newc{\mbmzmssmdrbar}{\ensuremath{ m_b(\mz)^{\drbar}_{{\mathrm{SUSY}}}}}
\newc{\mtaumzsmmsbar}{\ensuremath{ m_{\tau}(\mz)^{\msbar}_{{\mathrm{SM}}}}}
\newc{\mtaumzsmdrbar}{\ensuremath{ m_{\tau}(\mz)^{\drbar}_{{\mathrm{SM}}}}}
\newc{\mtaumzmssmdrbar}{\ensuremath{ m_{\tau}(\mz)^{\drbar}_{{\mathrm{SUSY}}}}}
\newc{\alphasmzms}{\ensuremath{\alpha_s(M_Z)^{\overline{MS}}}}
\newc{\alphaimzms}[1]{\ensuremath{\alpha_{#1}(M_Z)^{\overline{MS}}}}
\newc{\alphaemmz}{\ensuremath{\alpha_{\mathrm{em}}(M_Z)^{\overline{MS}}}}

\newc{\mzero}{\ensuremath{{m_0}}}
\newc{\mhalf}{\ensuremath{ m_{1/2}}}
\newc{\tanb}{\ensuremath{\tan\beta}}
\newc{\azero}{\ensuremath{ A_0}}
\newc{\bzero}{\ensuremath{ B_0}}
\newc{\signmu}{\ensuremath{\rm{sgn}\,\mu}}
\newc{\mueff}{\ensuremath{\mu_{\rm{eff}}}}
\newc{\lam}{\ensuremath{{\lambda}}}
\newc{\kap}{\ensuremath{{\kappa}}}
\newc{\alam}{\ensuremath{{A_{\lambda}}}}
\newc{\akap}{\ensuremath{{A_{\kappa}}}}
\newc{\hs}{\ensuremath{ H_s}}      
\newc{\mhs}{\ensuremath{ m_{H_s}}} 
\newc{\mgut}{\ensuremath{ M_{\rm GUT}}}
\newc{\mplanck}{\ensuremath{ M_{\rm P}}}      \newc{\mpl}{\ensuremath{ M_{\rm Pl}}}
\newc{\msusy}{\ensuremath{ M_{\rm SUSY}}}      \newc{\ms}{\ensuremath{ M_{\rm S}}}
 \newc{\mhl}{\ensuremath{m_\hl}} 
 \newc{\mhone}{\ensuremath{m_{h_1}}} 
 \newc{\mhtwo}{\ensuremath{m_{h_2}}} 
 \newc{\mglu}{\ensuremath{m_{\tilde g}}} 
 \newc{\mul}{\ensuremath{m_{\tilde{u}_L}}} 
 \newc{\mtone}{\ensuremath{m_{\tilde{t}_1}}} 
 \newc{\ma}{\ensuremath{m_A}} 
 \newc{\maone}{\ensuremath{m_{a_1}}} 
 \newc{\matwo}{\ensuremath{m_{a_2}}}
 \newc{\hone}{\ensuremath{h_1}}
 \newc{\htwo}{\ensuremath{h_2}}
 \newc{\aone}{\ensuremath{a_1}}
 \newc{\atwo}{\ensuremath{a_2}}
 \newc{\mhu}{\ensuremath{ m_{H_u}}}       
 \newc{\mhd}{\ensuremath{ m_{H_d}}}
 \newc{\mhusq}{\ensuremath{ m_{H_u}^2}}       
 \newc{\mhdsq}{\ensuremath{ m_{H_d}^2}}
 \newc{\mhuew}{\ensuremath{ m^{\ast}_{H_u}}}       
 \newc{\mhdew}{\ensuremath{ m^{\ast}_{H_d}}}
 \newc{\mhuewsq}{\ensuremath{ m^{\ast\, 2}_{H_u}}}       
 \newc{\mhdewsq}{\ensuremath{ m^{\ast\, 2}_{H_d}}}
 \newc{\hu}{\ensuremath{ H_u}}       
 \newc{\hd}{\ensuremath{ H_d}}
 \newc{\barmhu}{\ensuremath{ \bar{m}_{H_u}}}
 \newc{\barmhd}{\ensuremath{ \bar{m}_{H_d}}}

 \newc{\mqthree}{\ensuremath{m_{\widetilde{Q}_3}^2}}
 \newc{\muthree}{\ensuremath{m_{\tilde{u}_3}^2}}
 \newc{\mdthree}{\ensuremath{m_{\tilde{d}_3}^2}}
 \newc{\mlthree}{\ensuremath{m_{\widetilde{L}_3}^2}}
 \newc{\methree}{\ensuremath{m_{\tilde{e}_3}^2}}
 \newc{\mqtwo}{\ensuremath{m_{\widetilde{Q}_2}^2}}
 \newc{\mutwo}{\ensuremath{m_{\tilde{u}_2}^2}}
 \newc{\mdtwo}{\ensuremath{m_{\tilde{d}_2}^2}}
 \newc{\mltwo}{\ensuremath{m_{\widetilde{L}_2}^2}}
 \newc{\metwo}{\ensuremath{m_{\tilde{e}_2}^2}}
 \newc{\mqone}{\ensuremath{m_{\widetilde{Q}_1}^2}}
 \newc{\muone}{\ensuremath{m_{\tilde{u}_1}^2}}
 \newc{\mdone}{\ensuremath{m_{\tilde{d}_1}^2}}
 \newc{\mlone}{\ensuremath{m_{\widetilde{L}_1}^2}}
 \newc{\meone}{\ensuremath{m_{\tilde{e}_1}^2}}
 \newc{\msmul}{\ensuremath{m_{\tilde{\mu}_L}}}
 \newc{\msmur}{\ensuremath{m_{\tilde{\mu}_R}}}
 \newc{\msneumu}{\ensuremath{m_{\tilde{\nu}_{\mu}}}}
 \newc{\mone}{\ensuremath{M_1}}
 \newc{\monesq}{\ensuremath{M_1^2}}
 \newc{\mtwo}{\ensuremath{M_2}}
 \newc{\mtwosq}{\ensuremath{M_2^2}}
 \newc{\mthree}{\ensuremath{M_3}}
 \newc{\mthreesq}{\ensuremath{M_3^2}}
 \newc{\atau}{\ensuremath{{A_{\tau}}}}
 \newc{\at}{\ensuremath{{A_{t}}}}
 \newc{\ab}{\ensuremath{{A_{b}}}}
 \newc{\atausq}{\ensuremath{{A_{\tau}^2}}}
 \newc{\atsq}{\ensuremath{{A_{t}^2}}}
 \newc{\absq}{\ensuremath{{A_{b}^2}}}

 \newc{\dmzero}{\ensuremath{\Delta{_{m_0}}}}
 \newc{\dmhalf}{\ensuremath{\Delta{_{m_{1/2}}}}}
 \newc{\dmu}{\ensuremath{\Delta{_{\mu}}}}

 \newc{\pten}{\ensuremath{\psi_{10}}}
 \newc{\ffive}{\ensuremath{\phi_{5}}}
 \newc{\hfive}{\ensuremath{h_{5}}}
 \newc{\hbfive}{\ensuremath{h_{\bar{5}}}}
 \newc{\thet}{\ensuremath{\theta_{50}}}
 \newc{\thetb}{\ensuremath{\theta_{\,\overline{50}}}}
 \newc{\ptenhat}{\ensuremath{\hat{\psi}_{10}}}
 \newc{\ffivehat}{\ensuremath{\hat{\phi}_{5}}}
 \newc{\hfivehat}{\ensuremath{\hat{h}_{5}}}
 \newc{\hbfivehat}{\ensuremath{\hat{h}_{\bar{5}}}}
 \newc{\thethat}{\ensuremath{\hat{\theta}_{50}}}
 \newc{\thetbhat}{\ensuremath{\hat{\theta}_{\,\overline{50}}}}
 \newc{\si}{\ensuremath{\Sigma}}
 \newc{\mfive}{\ensuremath{m_5^2}}
 \newc{\mten}{\ensuremath{m_{10}^2}}
 \newc{\dfive}{\ensuremath{\Delta^2_5}}
 \newc{\dbfive}{\ensuremath{\Delta^2_{\bar{5}}}}
 \newc{\dfifty}{\ensuremath{\Delta^2_{50}}}
 \newc{\dfiftyb}{\ensuremath{\Delta^2_{\,\overline{50}}}}
 \newc{\msi}{\ensuremath{m_{\Sigma}^2}}
 \newc{\lamh}{\ensuremath{\lambda_{H}}}
 \newc{\lamhb}{\ensuremath{\lambda_{\bar{H}}}}
 \newc{\ah}{\ensuremath{A_{H}}}
 \newc{\ahb}{\ensuremath{A_{\bar{H}}}}
 \newc{\lams}{\ensuremath{\lambda_{S}}}
 \newc{\as}{\ensuremath{A_{S}}}
 \newc{\lamsig}{\ensuremath{\lambda_{\si}}}
 \newc{\asig}{\ensuremath{A_{\si}}}

 \newc{\msten}{\ensuremath{m_{16}^2}}
 \newc{\mhun}{\ensuremath{m_{126}^2}}
 \newc{\mhunb}{\ensuremath{m_{\bar{126}}^2}}
 \newc{\mthun}{\ensuremath{m_{210}^2}}
 \newc{\ahun}{\ensuremath{A_{\bar{126}}}}
 \newc{\yhun}{\ensuremath{Y_{\bar{126}}}}
 \newc{\aten}{\ensuremath{A_{10}}}
 \newc{\yten}{\ensuremath{Y_{10}}}
 \newc{\alone}{\ensuremath{A_{\lambda_1}}}
 \newc{\altwo}{\ensuremath{A_{\lambda_2}}}
 \newc{\althree}{\ensuremath{A_{\lambda_3}}}
 \newc{\althreeb}{\ensuremath{A_{\bar{\lambda_3}}}}
 \newc{\lone}{\ensuremath{\lambda_1}}
 \newc{\ltwo}{\ensuremath{\lambda_2}}
 \newc{\lthree}{\ensuremath{\lambda_3}}
 \newc{\lthreeb}{\ensuremath{\bar{\lambda_3}}}

\newc{\sigsip}{\ensuremath{\sigma^{\rm SI}_{p}}}	\newc{\sigsin}{\ensuremath{\sigma^{\rm SI}_{n}}}
\newc{\sigsdp}{\ensuremath{\sigma^{\rm SD}_{p}}}	\newc{\sigsdn}{\ensuremath{\sigma^{\rm SD}_{n}}}
\newc{\sigsi}{\ensuremath{\sigma^{\rm SI}}}	\newc{\sigsd}{\ensuremath{\sigma^{\rm SD}}}
\newc{\sigv}{\ensuremath{\sigma v}}
\newc{\abund}{\ensuremath{ \Omega h^2}}
\newc{\omegadm}{\ensuremath{ \Omega_{{\rm DM}}}}     \newc{\abunddm}{\ensuremath{ \Omega_{{\rm DM}} h^2}} 
\newc{\omegam}{\ensuremath{ \Omega_{{\rm m}}}}       \newc{\abundm}{\ensuremath{ \Omega_{{\rm m}} h^2}}
\newc{\omegab}{\ensuremath{ \Omega_{{\rm b}}}}	\newc{\abundb}{\ensuremath{ \Omega_{{\rm b}} h^2}}
\newc{\omegatot}{\ensuremath{ \Omega_{{\rm TOT}}}}
\newc{\omegacdm}{\ensuremath{ \Omega_{{\rm CDM}}}}   \newc{\abundcdm}{\ensuremath{ \Omega_{{\rm CDM}} h^2}}
\newc{\omegalambda}{\ensuremath{ \Omega_{\Lambda}}} \newc{\abundlambda}{\ensuremath{ \Omega_{\Lambda} h^2}}
\newc{\omegarad}{\ensuremath{ \Omega_{{\rm rad}}}}  \newc{\abundrad}{\ensuremath{ \Omega_{{\rm rad}} h^2}}
\newc{\rhocrit}{\ensuremath{ \rho_{\rm crit}}}
\newc{\rhochi}{\ensuremath{ \rho_{\chi}}}
\newc{\abunchi}{\ensuremath{\Omega_\chi h^2}}
\newc{\abundlsp}{\ensuremath{\Omega_{\rm LSP}h^2}}
\newcommand*{\abundchi}{\ensuremath{\Omega_\chi h^2}}

\newc{\amu}{\ensuremath{ a_{\mu}}}        \newc{\amususy}{\ensuremath{ a_{\mu}^{\mathrm{SUSY}}}}
\newc{\amuexpt}{\ensuremath{ a_{\mu}^{\mathrm{expt}}}}        \newc{\amusm}{\ensuremath{ a_{\mu}^{\mathrm{SM}}}}
\newc\deltaamu{\ensuremath{\Delta a_{\mu}}} \newc{\deltaamususy}{\ensuremath{\delta a_{\mu}^{\mathrm{SUSY}}}}
\newc\gmtwo{\ensuremath{ (g-2)_{\mu}}} 
\newc{\deltagmtwomususy}{\ensuremath{\delta\left(g-2\right)_{\mu}^{\mathrm{SUSY}}}}
\newc{\deltagmtwomu}{\ensuremath{\delta\left(g-2\right)_{\mu}}}
\newc\BR{\ensuremath{\rm BR}}
\newc\bsgamma{\ensuremath{ b\rightarrow s \gamma }}
\newc\bxsgamma{\ensuremath{\overline{B}\rightarrow X_{s}\gamma}}
\newc\brbsgamma{\ensuremath{\BR\left(\bsgamma\right)}}
\newc\brbxsgamma{\ensuremath{\BR\left(\bxsgamma\right)}}
\newc\bsmumu{\ensuremath{B_s\to\mu^+\mu^-}}
\newc\brbsmumu{\ensuremath{\BR\left(B_s\to\mu^+\mu^-\right)}}
\newc\bdmmumu{\ensuremath{\overline{B}_d\to\mu^+\mu^-}}
\newc\bbbarmix{\ensuremath{\overline{B}_s\mbox{-}B_s}}      
\newc\delmbs{\ensuremath{\Delta M_{B_s}}}
\newc{\butaunu}{\ensuremath{B_u \rightarrow \tau \nu}}
\newc{\brbutaunu}{\ensuremath{\BR\left(B_u \rightarrow \tau \nu\right)}}

\newcommand*{\reftable}[1]{Table~\ref{#1}}         
       
\newcommand*{\reffig}[1]{Fig.~\ref{#1}}

     \newcommand*{\refsec}[1]{Sec.~\ref{#1}}

\newcommand*{\neutone}{\ensuremath{\tilde{\chi}^0_1}}
\newcommand*{\neuttwo}{\ensuremath{\tilde{{\chi}}^0_2}}

\newcommand*{\charone}{\ensuremath{\chi^{\pm}_1}}

\newcommand*{\stau}{\ensuremath{\tilde{\tau}}}
\newcommand*{\slep}{\ensuremath{\tilde{l}}}

\newcommand*{\mstopone}{\ensuremath{m_{\tilde{t}_1}}}

\newcommand*{\pythia}{\text{PYTHIA}}

\newcommand*{\checkm}{\text{CheckMATE}}
\newcommand*{\madgr}{\texttt{MadGraph5\_aMC@NLO}}
\newcommand*{\delphes}{\text{DELPHES 3}}

\let\oldcite\cite
\renewcommand*{\cite}{~\oldcite}

\newcommand*{\hl}{\ensuremath{h}}

\restylefloat{figure}

\title{GUT-inspired SUSY and the muon $g-2$ anomaly: prospects for LHC 14 TeV}

\author{Kamila Kowalska,}
\author[1]{Leszek Roszkowski,\note{On leave of absence from the University of Sheffield, U.K.}}
\author{Enrico Maria Sessolo}
\author{and Andrew J.~Williams}

\affiliation{National Centre for Nuclear Research,\\
  Ho{\. z}a 69, 00-681 Warsaw, Poland} 

\emailAdd{Kamila.Kowalska@fuw.edu.pl}
\emailAdd{L.Roszkowski@sheffield.ac.uk}
\emailAdd{Enrico-Maria.Sessolo@fuw.edu.pl}
\emailAdd{Andrew.Williams@fuw.edu.pl}

\abstract{We consider the possibility that the muon $g-2$ anomaly, \deltagmtwomu, finds its origins in low energy supersymmetry (SUSY).
In the general MSSM the parameter space consistent with \deltagmtwomu\ and correct dark matter relic density of the lightest
neutralino easily evades the present direct LHC limits on sparticle masses and also lies to a large extent beyond future LHC sensitivity.
The situation is quite different in GUT-defined scenarios where input SUSY parameters are no longer independent.
We analyze to what extent the LHC can probe a broad class of GUT-inspired SUSY models with gaugino non-universality
that are currently in agreement with the bounds from \deltagmtwomu, as well as with the relic density and the Higgs mass measurement.
To this end we perform a detailed numerical simulation of several searches for electroweakino and slepton production at the LHC and
derive projections for the LHC 14\tev\ run.
We show that, within GUT-scale SUSY there is still plenty of room for the explanation of the muon anomaly, 
although the current LHC data already imply strong limits on the parameter space consistent with \deltagmtwomu.
On the other hand, we demonstrate that the parameter space will be basically fully explored within the sensitivity of the 14\tev\ run with 300\invfb.
This opens up the interesting possibility that, if the \gmtwo\ anomaly is real
then some positive signals must be detected at the LHC, or else these models will be essentially ruled out.
Finally, we identify the few surviving spectra that will provide a challenge
for detection at the LHC 14\tev\ run and we characterize their properties.}

\begin{document}
\maketitle

\section{\label{sec:intro}Introduction}

The measurement of the anomalous magnetic moment of the muon, \gmtwo, by the Brookhaven experiment\cite{Bennett:2006fi} a decade ago shows a $\sim 3\sigma$ discrepancy with the Standard Model (SM). 
The measured value of this discrepancy,  $\deltagmtwomu=a_{\mu}^{\textrm{exp}}-a_{\mu}^{\textrm{SM}}$, is
\bea
\deltagmtwomu & = & (28.7\pm 8.0)\times 10^{-10}\,\textrm{ or}\label{measure1}\\
\deltagmtwomu & = & (26.1\pm 8.0)\times 10^{-10}\,,\label{measure2}
\eea
according to whether the lowest order hadronic contributions from 
Ref.\cite{Davier:2010nc} or Ref.\cite{Hagiwara:2011af} are used to compute the SM value. 
The discrepancy will soon be either confirmed or overruled by the New Muon g-2 experiment at Fermilab\cite{Roberts:2010cj,Grange:2015eea}, 
which is bound to rekindle the interest of the particle physics community in the subject. 

A $\sim 3\sigma$ difference with the SM can easily be accommodated in the Minimal Supersymmetric Standard Model (MSSM)\cite{Czarnecki:2001pv,Martin:2001st,Feng:2001tr,Everett:2001tq,Chattopadhyay:2001vx,
Komine:2001fz,Ellis:2001yu,Arnowitt:2001be,Baer:2001kn,Baltz:2001ts,Fukuyama:2003hn,Stockinger:2006zn,Cho:2011rk,
Endo:2013bba,Mohanty:2013soa,Ibe:2013oha,Akula:2013ioa,Fowlie:2013oua,
Endo:2013lva,Freitas:2014pua,Ajaib:2014ana,Gogoladze:2014cha,Chakraborti:2014gea,Das:2014kwa,
Li:2014dna,Ajaib:2015ika,Calibbi:2015kja}, 
which is rich in particles with the right couplings to provide significant loop contributions to the $\mu\mu\gamma$ vertex.
The 1-loop contributions in the MSSM\cite{Moroi:1995yh,Martin:2001st} are roughly split into those arising from   
a chargino/sneutrino loop and those given by smuon/neutralino loops, so that 
at the lowest order the measurement of \deltagmtwomu\ can be parametrized by
\be
\mu, \mone, \mtwo, \msmul, \msmur, \msneumu, \tanb\,,\label{g2pars} 
\ee
where $\mu$ is the higgsino mass parameter, \tanb\ is the ratio of the Higgs vacuum expectation values (vev's), 
\mone\ and \mtwo\ are the bino and wino soft supersymmetry-breaking masses,
\msmul\ and \msmur\ are smuon soft masses, and \msneumu\ is the soft mass of the muon sneutrino.  

The LHC has started to test the electroweak (EW) sector of the MSSM in a class of searches involving different multiplicities of leptons, no jets, 
and a significant amount of missing energy. The 8\tev\ run has provided strong lower bounds on chargino and slepton masses, 
in particular when interpreted in the framework of simplified model spectra (SMS)\cite{Chatrchyan:2013sza}. 
However, several studies have  shown\cite{Endo:2013bba,Akula:2013ioa,Fowlie:2013oua,Chakraborti:2014gea,Das:2014kwa} that if the experimental 
limits provided by the CMS and ATLAS collaborations are reinterpreted and applied to more general MSSM scenarios 
the 8\tev\ LHC results can only constrain a small part of the available parameter space, so that ample room still remains to attribute a 
supersymmetric (SUSY) origin to \deltagmtwomu.

On the other hand, in scenarios where unification of the scalar and gaugino masses is imposed as a remnant of some 
new physics at the scale of Grand Unification (GUT scale), 
like in the well-known Constrained MSSM (CMSSM) or the Non-Universal Higgs Mass (NUHM) model, 
it has become virtually impossible to find regions of the parameter space where 
the measurement of \deltagmtwomu\ can be reproduced 
(see, e.g.,\cite{Bechtle:2012zk,Fowlie:2012im,Buchmueller:2012hv,Strege:2012bt}). 
The reason is well known, and is due to the combined effect of direct lower limits on colored sparticles at the LHC
and the discovery of a Higgs boson with $\mhl\simeq 125\gev$\cite{Aad:2012tfa,Chatrchyan:2012ufa,Aad:2015zhl}, 
which together have pushed the favored parameter space for unified scalar and 
gaugino masses to the multi-\tev\ regime, thus indirectly forbidding the possibility of light sleptons, binos, and winos.   

Interestingly, this is not necessarily the case for GUT-scale models in which the assumption of gaugino unification is relaxed. 
It has been pointed out, e.g., in Refs.\cite{Akula:2013ioa,Nath:2015dza} 
(but see also\cite{Mohanty:2013soa,Chakrabortty:2013voa,Gogoladze:2014cha,Ajaib:2015ika}), that in models of gravity 
mediation all one really needs is GUT-defined boundary conditions such that the high-scale value of the gluino soft mass, $M_3$, is much
larger than the values of \mone\ and \mtwo. Through the renormalization group equations (RGEs) large initial conditions for $M_3$ 
can then drive the masses of sparticles charged under $SU(3)$ to large values at the EW scale, 
in agreement with the LHC data and the measurement of the Higgs boson, 
while the sparticles charged under only the EW gauge groups remain quite light, in agreement with the measurement of \deltagmtwomu. 

The goal of this paper is to examine to what extent the oncoming run of the LHC
can probe the parameter space of these gravity-mediated, GUT-scale SUSY models that satisfy 
the present constraints for \gmtwo. Besides being motivated by considerations of symmetry and providing a realistic framework for SUSY breaking, 
these scenarios are more constrained than generic phenomenological parametrizations of the MSSM 
by the measurement of the relic abundance 
of dark matter (DM) at Planck or WMAP\cite{Ade:2013zuv,Komatsu:2010fb}, $\abund\simeq 0.12$.
Once this bound is combined with the measurement of \deltagmtwomu, the allowed parameter space becomes significantly reduced.
For example, we shall see that often the lightest SUSY particle (LSP) is a fairly light bino-like neutralino $\chi$, 
which needs an equally light slepton or, alternatively, substantial mixing with higgsinos to yield $\abundchi\lesssim 0.12$.
Thus, the parameter space that gives \abundchi\ and \deltagmtwomu\ is in this case  
particularly sensitive to direct LHC searches involving the production and decay of sleptons and electroweakinos.

In this paper we reinterpret existing 3-lepton\cite{Aad:2014nua,Khachatryan:2014qwa} and 2-lepton\cite{Aad:2014vma} 
LHC searches for direct production of charginos, neutralinos, and sleptons, and apply them to the allowed parameter space 
of several GUT-scale SUSY models characterized by non-universal boundary conditions to fit the \gmtwo\ measurement.
We progressively increase the complexity of the analyzed models by disunifying several parameters defined at the GUT scale. 
We do this by following patterns typical of GUT symmetries like $SO(10)$, $SU(5)$ or Pati-Salam.
From the phenomenological point of view, this is 
equivalent to introducing an increasing number of mechanisms that yield the correct relic abundance of the LSP. 

We will show that the present constraints from the LHC on the EW sector of this 
large class of models
are already quite stringent and, more importantly, that the 14\tev\ run offers the opportunity 
to probe the remaining parameter space virtually in its entirety.
To this end, we calculate the projected sensitivity of 2- and 3-lepton searches for the LHC 14\tev\ run with 300\invfb\
and apply the simulations to the defined GUT-scale models. 
Incidentally, our projections can be compared for SMS scenarios to the ones produced by other
groups\cite{Eckel:2014dza}. 

The paper is organized as follows. 
In \refsec{sec:g2andrelic} we review the interplay of the bounds 
from the measurement of \deltagmtwomu\ and \abundchi\ on the parameter space of the MSSM,
and we will comment on the LHC reach in the phenomenological MSSM. 
In \refsec{sec:scenarios} we introduce the GUT-scale models that will be scanned over 
and expose the parameter space consistent with several phenomenological constraints. 
In \refsec{sec:results} we present our methodology for reinterpretation of LHC searches 
and our projections for the next run of the LHC in SMS scenarios.
We then show the main results, i.e., we apply the derived 
LHC bounds and projections to the GUT scenarios defined in \refsec{sec:scenarios}.
We finally present our summary and conclusions in \refsec{sec:conclusions}. 
   
\section{The relic density and $\mathbf{\gmtwo}$ in the MSSM\label{sec:g2andrelic}}

We review in this section the relations that can be derived on the parameter space of the MSSM when the 
measurement of \deltagmtwomu\ is considered in combination with the relic density.
A study on the subject that includes the limits from the LHC 8\tev\ run is done, e.g., in Ref.\cite{Chakraborti:2014gea}.
Our goal here is to show that these relations can be very useful in deriving bounds,
but unless some assumption about the mechanism of SUSY breaking is made,
there remain large fractions of the parameter space outside of the LHC reach.

The MSSM contributions to \deltagmtwomu\ have been calculated at 1 loop 
in\cite{Kosower:1983yw,Yuan:1984ww,Romao:1984pn,Vendramin:1988rd,Moroi:1995yh,Martin:2001st} using the mass insertion 
method.\footnote{A code to calculate the 1-loop contributions to \deltagmtwomu\ in generic new-physics models has recently become available\cite{Queiroz:2014zfa}.} 
Higher order contributions have been computed 
in\cite{Degrassi:1998es,Chang:2000ii,Chen:2001kn,Arhrib:2001xx,Heinemeyer:2003dq,Heinemeyer:2004yq,Feng:2006ei,Fargnoli:2013zda,Fargnoli:2013zia}.
At 1 loop there are five main contributions that can be split into two classes of diagrams:
chargino/sneutrino and neutralino/smuon contributions.
In terms of the MSSM parameters these five contributions are given by\cite{Moroi:1995yh}: 
\be 
\Delta_{\charone \tilde{\nu}_{\mu}}=\frac{g^2}{(4 \pi)^2}\frac{m_{\mu}^2\tanb}{\mu\mtwo}\,\mathcal{F}_{[\charone \tilde{\nu}_{\mu}]}\left(\frac{\mu^2}{\msneumu^2},\frac{\mtwo^2}{\msneumu^2}\right)\,,\label{charsneu}
\ee
\be
\Delta^{(1)}_{\chi\, \tilde{\mu}}=-\frac{1}{2}\frac{g^2}{(4 \pi)^2}\frac{m_{\mu}^2\tanb}{\mu\mtwo}\,\mathcal{F}_{[\chi\, \tilde{\mu}]}\left(\frac{\mu^2}{\msmul^2},\frac{\mtwo^2}{\msmul^2}\right)\,,
\ee
\be
\Delta^{(2)}_{\chi\, \tilde{\mu}}=\frac{1}{2}\frac{g'^2}{(4 \pi)^2}\frac{m_{\mu}^2\tanb}{\mu\mone}\,\mathcal{F}_{[\chi\, \tilde{\mu}]}\left(\frac{\mu^2}{\msmul^2},\frac{\mone^2}{\msmul^2}\right)\label{negl1}\,,
\ee
\be
\Delta^{(3)}_{\chi\, \tilde{\mu}}=-\frac{g'^2}{(4 \pi)^2}\frac{m_{\mu}^2\tanb}{\mu\mone}\,\mathcal{F}_{[\chi\, \tilde{\mu}]}\left(\frac{\mu^2}{\msmur^2},\frac{\mone^2}{\msmur^2}\right)\label{negl2}\,,
\ee
\be
\Delta^{(4)}_{\chi\, \tilde{\mu}}=\frac{g'^2}{(4 \pi)^2}\frac{m_{\mu}^2\mone\mu}{\msmul^2 \msmur^2}\tanb\,\mathcal{F}_{[\chi\, \tilde{\mu}]}\left(\frac{\msmur^2}{\mone^2},\frac{\msmul^2}{\mone^2}\right)\,,\label{neutsmu}
\ee
where $g$ and $g'$ are the gauge couplings of the $SU(2)$ and $U(1)$ SM groups, respectively, and the $\mathcal{F}_{[\charone \tilde{\nu}_{\mu}]}$ and $\mathcal{F}_{[\chi\, \tilde{\mu}]}$ are loop functions
that read
\bea 
\mathcal{F}_{[\charone \tilde{\nu}_{\mu}]}(x,y)&=& xy\left\{\frac{5-3(x+y)+xy}{(x-1)^2(y-1)^2}-\frac{2}{x-y}\left[\frac{\ln x}{(x-1)^3}-\frac{\ln y}{(y-1)^3}\right]\right\}\,,\\
\mathcal{F}_{[\chi\,\tilde{\mu}]}(x,y)&=& xy\left\{\frac{-3+x+y+xy}{(x-1)^2(y-1)^2}+\frac{2}{x-y}\left[\frac{x\ln x}{(x-1)^3}-\frac{y\ln y}{(y-1)^3}\right]\right\}\,,
\eea
where we have used the reduced forms of Ref.\cite{Endo:2013bba}.
Note that the numerical coefficient in front of Eqs.~(\ref{negl1}) and (\ref{negl2}) depends on $g'^2$ so that these contributions
are in general suppressed with respect to Eq.~(\ref{charsneu}). 
The neutralino/smuon contribution of Eq.~(\ref{neutsmu}), however, depends directly on $\mu$.
When $\mu\gg M_1,\msmul,\msmur$ it can become the dominant one.
\medskip

%
%

\begin{table}[t]
   \centering
   \begin{tabular}{|c|c|c|c|c|c|c|} 
      \hline
      Region & \mone\ & \mtwo\ & $\mu$ & $m_{\slep_L}$ & $m_{\slep_R}$ \\
      \hline
      \footnotesize $Z$-funnel & $\approx M_Z/2$ & -- & $100-500\gev$ & $\lesssim 1.5\tev$ & -- \\
      \hline
      \footnotesize $h$-funnel & $\approx m_h/2$ & -- & $100-1000\gev$ & $\lesssim 1.5\tev$ & -- \\
      \hline
      \footnotesize Well-tempered & $100-700\gev$ & $>\mu$ & $\approx M_1$ & $\lesssim 1.5\tev$ & -- \\
      \hline
      \footnotesize $\slep_R$-coannihilation & $100-500\gev$ & -- & -- & $\lesssim 2-2.5\tev$ & $\approx M_1$ \\
      \hline
      \footnotesize $\slep_L$-coannihilation & $100-500\gev$ & -- & -- & $\approx M_1$ & $2-2.5\tev$ \\
      \hline
      \footnotesize $\stau$-coannihilation & $100-400\gev $ & -- & -- & $\lesssim 1\tev$ & $\lesssim 1\tev$ \\
      \hline
      \footnotesize Pure higgsino & -- & $>\mu$ & $\lesssim 600\gev$ & $\lesssim 1.5\tev$ & -- \\
      \hline
     \footnotesize Pure wino & $> \mtwo$ & $\lesssim 800\gev$ & -- & $\lesssim 1.5\tev$ & -- \\
      \hline
   \end{tabular}
   \caption{\footnotesize Regions in the MSSM giving \deltagmtwomu\ (1-loop) at $2\sigma$ and $\abundchi\lesssim 0.12$.
   The symbol $>$ means here ``greater but not \textit{orders of magnitude} greater than...," see Figs.~(\ref{fig:Res})-(\ref{fig:pures}).}
   \label{tab:regions}
\end{table}

The bounds on the parameters (\ref{g2pars}) that arise from the measurement of \deltagmtwomu\ 
can be combined with the bounds that come from imposing 
$\abundchi\lesssim0.12$ on the same regions of the parameter space. 
We review here the regions consistent with Eq.~(\ref{measure1}) at least at the $2\sigma$ 
level and show the correspondingly allowed parameter space.

We adopt in this section simplifying assumptions typical of many phenomenological parametrizations of the MSSM\cite{Djouadi:1998di}.
The soft SUSY-breaking parameters are defined at the EW scale and we assume 
that first and second generation slepton soft masses are degenerate
($m_{\tilde{e}_{L}}=m_{\tilde{\mu}_{L}}$ $=m_{\tilde{\nu}_e}=m_{\tilde{\nu}_{\mu}}\equiv m_{\tilde{l}_{L}}$ and similar identities apply to
right-handed sleptons).
The bounds obtained using the 1-loop calculation, Eqs.~(\ref{charsneu})-(\ref{neutsmu}), are not far off 
from the ones given by more precise higher-order calculations, so that for the semi-quantitative discussion of this section
we will limit ourselves to the former approximation.
In the numerical analysis that we present in the following sections 
we will consider higher-order contributions with the help of the latest numerical codes.

The bounds arising from the combination of \gmtwo\ and \abundchi\ are summarized in \reftable{tab:regions}
and discussed below.\medskip 

\noindent \textbf{$Z/h$-resonance}. 
As is well known, when the lightest neutralino (hereafter simply ``the neutralino" or  $\chi$) is in the mass range 
$30\gev\lesssim m_{\chi}\lesssim 62\gev$ the cross section for pair annihilation can be enhanced 
by the resonance with the $Z$ boson or the Higgs: $\chi\chi\rightarrow Z/h\rightarrow \textrm{SM}\,\textrm{SM}$\cite{Griest:1988ma,Ellis:1989pg}.
The neutralino is in this case predominantly bino-like. 

To undergo $Z$-resonance annihilation, when $\mchi\approx M_Z/2$, the neutralino must have a non-negligible 
higgsino component to maximize the coupling to the $Z$ boson, 
so that the cross section is given by, up to proportionality constants and phase-space integration,\cite{Griest:1988ma,Drees:1992am}
\be 
\sigma v\sim  \frac{g'^4}{\mchi^2\left(1-\frac{\mu^2}{M_1^2}\right)^2}\cdot\frac{1}{\left(4-\frac{M_Z^2}{\mchi^2}\right)^2+\left(\frac{\Gamma_Z M_Z}{\mchi^2}\right)^2}\,,\label{zres}
\ee   
where $\Gamma_Z\simeq2.5\gev$ is the $Z$ width. 
Obviously, when $\mu$ becomes large the cross section decreases and the relic density exceeds the measured value.
This effectively sets an upper bound, $\mu\lesssim 500\gev$ corresponding to $\abundchi\simeq 0.12$.
When $\mtwo>\mu$ one derives an upper bound on the mass of the (mostly higgsino-like) lightest chargino, 
$m_{\charone}\lesssim 500\gev$. 

The Higgs resonance\cite{Ellis:1989pg} is qualitatively similar to the $Z$-resonance with a few differences. 
The width of the Higgs boson is much narrower than the $Z$'s: $\Gamma_h\simeq 4\mev$\cite{Heinemeyer:2013tqa}, so that 
the cross section is more sensitive to the neutralino being on or off the resonance. 
But, more importantly, the cross section scales as $1/(1-\mu/\mone)^2$\cite{Drees:1992am} rather than
$1/(1-\mu^2/\mone^2)^2$ as in Eq.~(\ref{zres}), so that $\mu$ is less constrained than in the $Z$-resonance region,
$\mu\lesssim 1000\gev$ in the $h$-resonance region. On the other hand,
$Z$- and $h$-resonance regions both depend minimally on \tanb\ or the slepton masses, 
so that these parameters are not bounded by the relic density constraint.
The same is true for the wino soft mass, $M_2$, that can assume arbitrarily large values.

Because of the upper bound on $\mu$, in the $Z$ and $h$-resonance regions 
the dominant contribution to \deltagmtwomu\ is given by the chargino-sneutrino loop, Eq.~(\ref{charsneu}).
The parameter space allowed at $2\sigma$ by Eq.~(\ref{measure1}) strongly depends on the value of \tanb,
which cannot be constrained by \abundchi. 

\begin{figure}[t]
\centering
\subfloat[]{%
\label{fig:a}%
\includegraphics[width=0.47\textwidth]{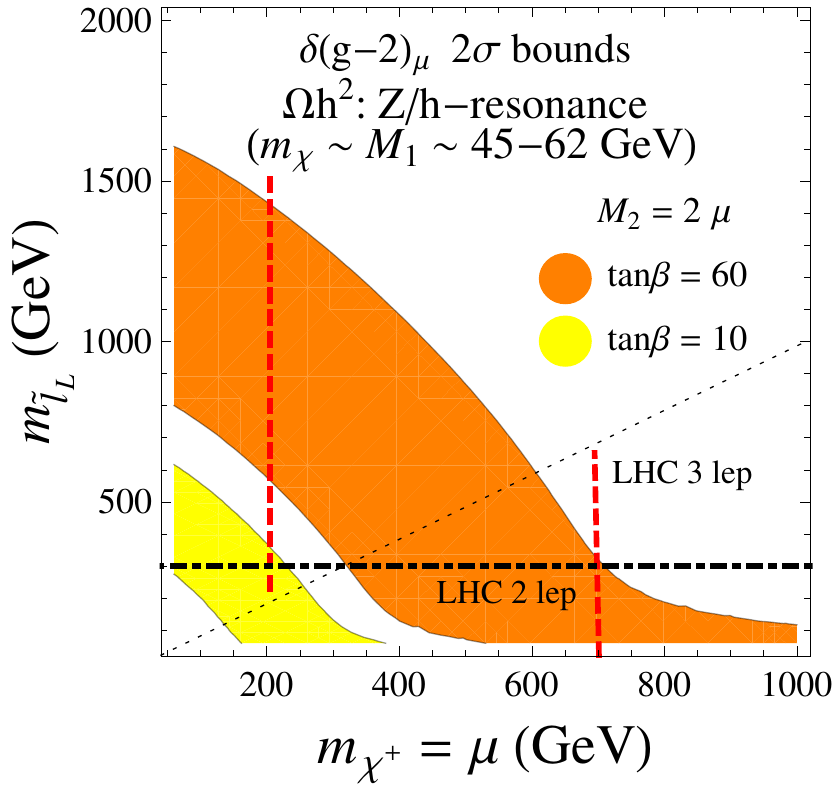}
}%
\hspace{0.02\textwidth}
\subfloat[]{%
\label{fig:b}%
\includegraphics[width=0.47\textwidth]{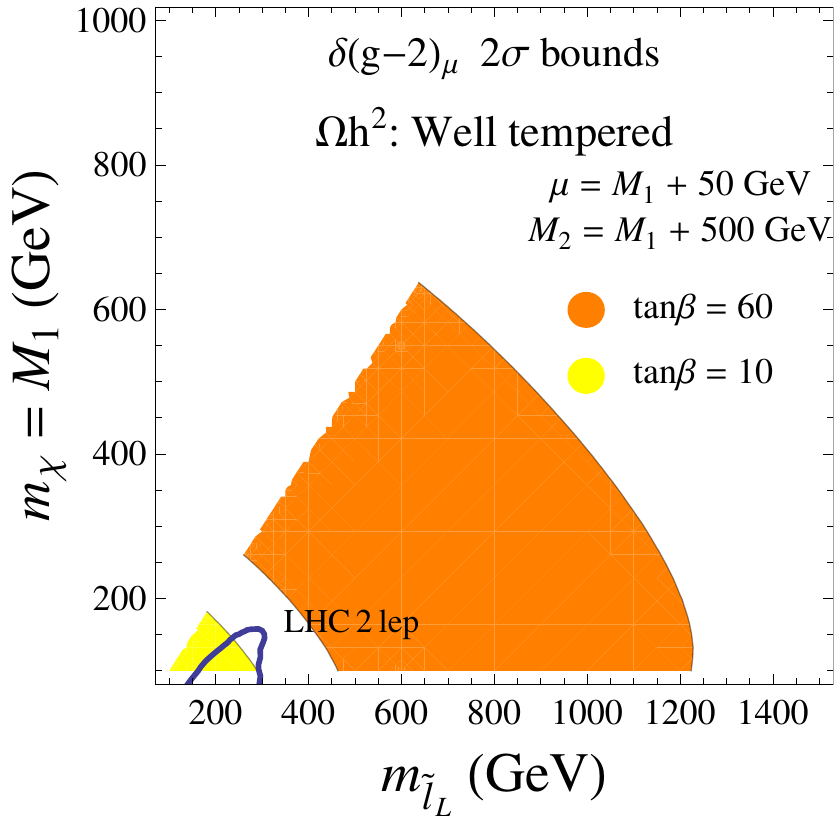}
}%
\caption{\footnotesize \protect\subref{fig:a} The parameter space belonging to the $Z/h$-resonance region of the MSSM that is 
allowed at $2\sigma$ by \gmtwo\ in the ($m_{\charone}$, $m_{\slep_L}$) plane. 
The orange band shows the case with $\tanb=60$ and the yellow band the one with $\tanb=10$. The black dot-dashed horizontal line
shows the approximate 95\%~C.L. lower bound for $m_{\slep_L}$ from 2 lepton searches\cite{Aad:2014vma,Khachatryan:2014qwa} 
at the LHC. The red dashed line shows the approximate lower bound in $m_{\charone}$ from 3 lepton searches\cite{Aad:2014nua,Khachatryan:2014qwa}, 
which differs if $m_{\slep_L} < m_{\charone}$ or viceversa.       
\protect\subref{fig:b} The allowed $2\sigma$ parameter space in the ($m_{\slep_L}$, \mchi) plane 
for the parameter space belonging to the region of mixed bino/higgsino neutralinos.
The solid dark blue line shows the approximate 95\%~C.L. exclusion bound from 2-lepton searches. The color code is the same as in \protect\subref{fig:a}.}  
\label{fig:Res}
\end{figure}

In \reffig{fig:Res}\subref{fig:a} we show in the ($m_{\charone}$, $m_{\slep_L}$) plane the part of the $Z/h$-resonance 
region consistent at $2\sigma$ with Eq.~(\ref{measure1}).
The colored bands show the \gmtwo\ bounds for two values of \tanb, $\tanb=60$ in orange and $\tanb=10$ in yellow.
The right-handed slepton mass is set to a large value, $m_{\slep_R}= 5\tev$, as 
the dominant contribution to \deltagmtwomu\ in this case, Eq.~(\ref{charsneu}),
does not depend on $m_{\slep_R}$. 
The plot shows that soft mass $m_{\slep_L}$ is bound to be lighter than 
$\sim 1.5\tev$ for $\tanb\simeq 60$,
but the upper bound becomes more restrictive as one considers smaller \tanb\ values. 
We assume here $\mtwo=2\mu$. For larger values of \mtwo\ the plot gets 
slowly squashed down, and at $\mtwo\simeq5\tev$ the limits on $m_{\slep_L}$
become approximately 3 times stronger.

It has been shown\cite{Endo:2013lva,Chakraborti:2014gea,Calibbi:2014lga} that the parameter space corresponding to the $Z/h$-resonance 
region can be probed at the LHC by 3-lepton searches for EW-ino production and 2-lepton searches for direct slepton production.
The bounds from 3-lepton searches are much stronger in the presence 
of an intermediate slepton between the mass of the chargino and neutralino\cite{Aad:2014nua,Khachatryan:2014qwa}.  
They are approximately shown as red dashed lines in \reffig{fig:Res}\subref{fig:a}.
The approximate bound from 2-lepton searches\cite{Aad:2014vma,Khachatryan:2014qwa} for slepton pair production
is shown as a dot-dashed black line.
One can see that for large \tanb\ a large fraction of the parameter space is presently not excluded. 
However, we will show in \refsec{sec:results} that the outlook for the 14\tev\ run improves considerably.\medskip

\noindent \textbf{Neutralino of mixed bino/higgsino composition}.
As one considers larger masses, for a bino-like neutralino it becomes necessary 
to increase the mixing with higgsino states to enhance the annihilation cross section. 
These ``well tempered" neutralinos\cite{ArkaniHamed:2006mb} efficiently annihilate to gauge bosons through $t$-channel chargino exchange.

In \reffig{fig:Res}\subref{fig:b} we show in the ($m_{\slep_L}$, \mchi) plane the region of the parameter space consistent 
at $2\sigma$ with Eq.~(\ref{measure1}) in the case of a mixed bino/higgsino neutralino with $\mchi\approx\mone\lesssim \mu$.
The colored bands show two different \tanb\ cases.
Again the right-handed slepton mass has been set at 5\tev, and the wino soft mass is set here to $\mtwo=\mone+500\gev$.
Raising \mtwo\ moves the allowed bounds down and left, to smaller values of \mchi\ and $m_{\slep_L}$, 
by reducing the contribution of Eq.~(\ref{charsneu}) to \deltagmtwomu.
Thus, \mtwo\ cannot be heavier than a few \tev.
The \gmtwo\ constraint requires approximately, $m_{\slep_L}\lesssim 1.2-1.5\tev$ and $\mchi\lesssim 700\gev$.

This region is notoriously difficult to probe in 3-lepton searches at the LHC, 
because the masses of the lightest chargino and neutralino are almost degenerate.
On the other hand, the bounds from 2-lepton searches, approximately indicated with a solid dark blue line, are 
much too weak at the moment to probe the parameter space.\medskip

\noindent \textbf{Neutralino/slepton coannihilation}.
For a predominantly bino-like neutralino the correct relic density can be obtained 
if $\chi$ coannihilates with an almost degenerate slepton\cite{Griest:1990kh,Ellis:1998kh}. 
This mechanism is particularly important for what follows, as it is one of the few realized in 
models with GUT-scale boundary conditions.

In most cases the neutralino mass must be within $\sim 20\gev$ of the mass of the coannihilating 
slepton, and it also becomes very hard to compensate for the increasing mass when 
\mchi\ reaches an approximate upper bound of $\sim 500\gev$.    

\begin{figure}[t]
\centering
\subfloat[]{%
\label{fig:a}%
\includegraphics[width=0.47\textwidth]{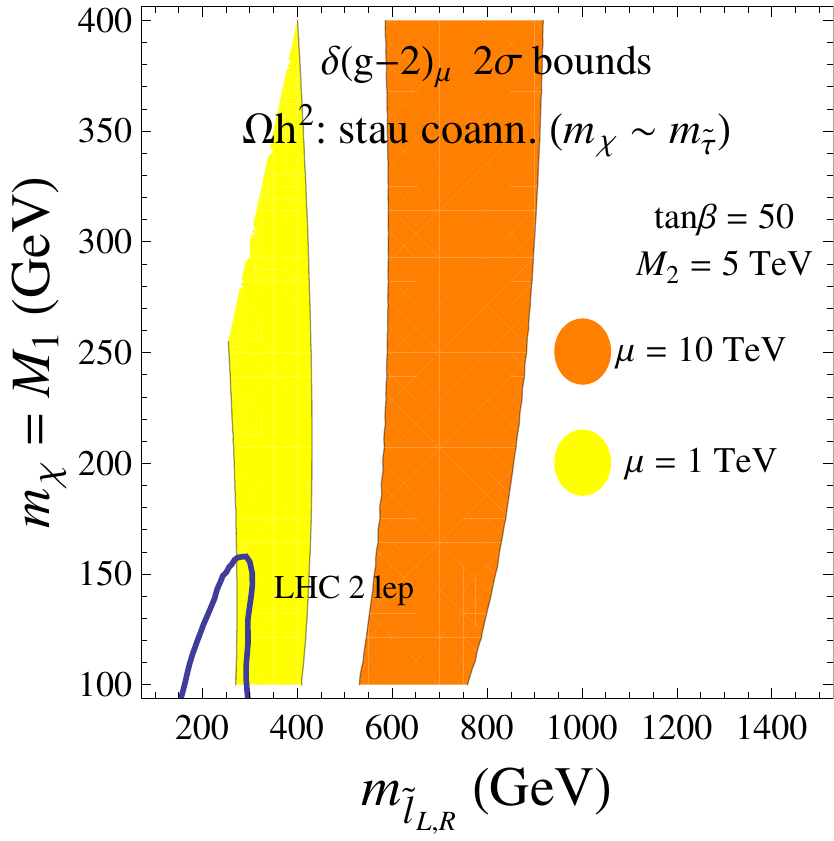}
}%
\hspace{0.02\textwidth}
\subfloat[]{%
\label{fig:b}%
\includegraphics[width=0.47\textwidth]{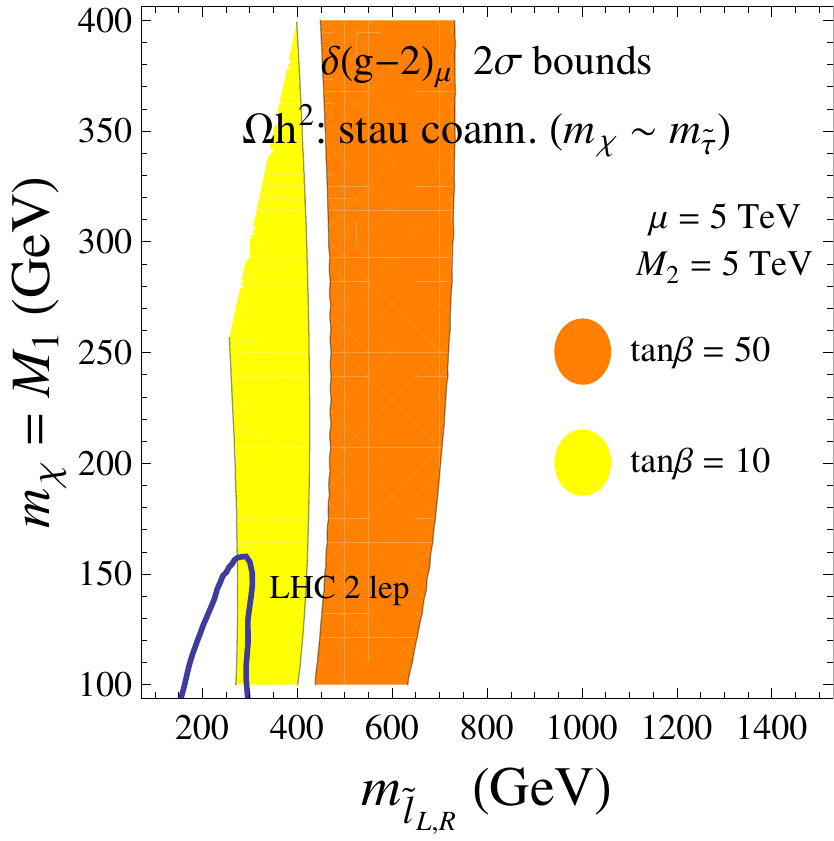}
}%

\subfloat[]{%
\label{fig:c}%
\includegraphics[width=0.47\textwidth]{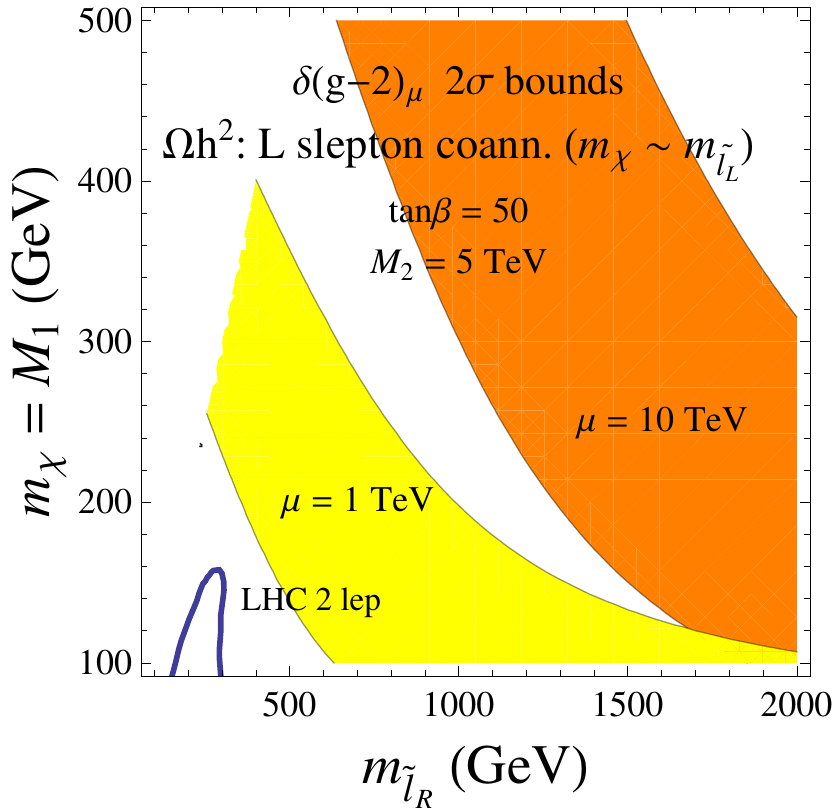}
}%
\hspace{0.02\textwidth}
\subfloat[]{%
\label{fig:d}%
\includegraphics[width=0.47\textwidth]{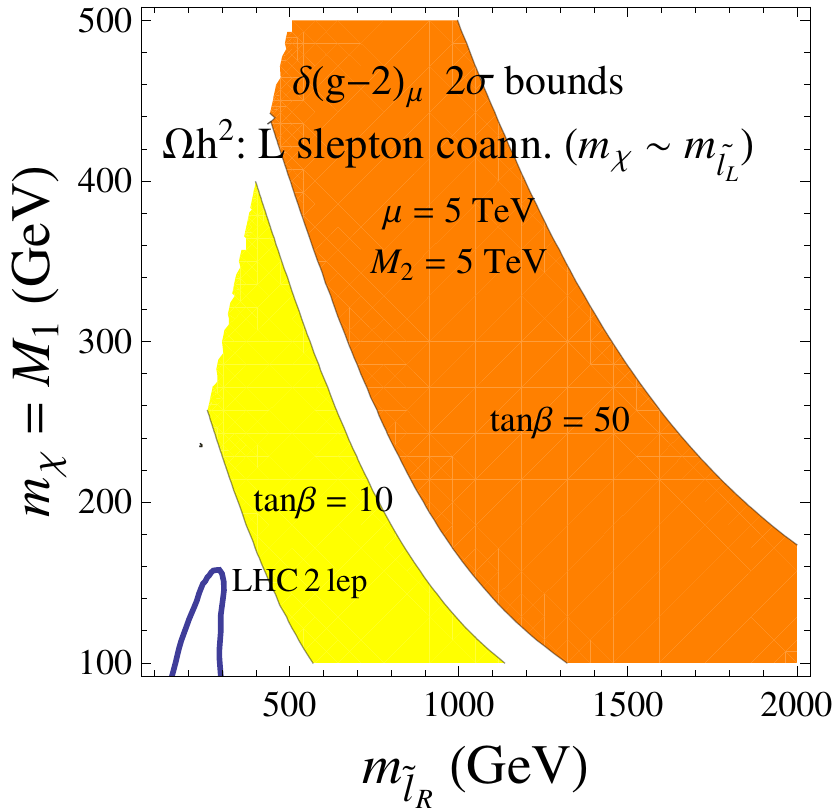}
}%
\caption{\footnotesize \protect\subref{fig:a} The parameter space allowed at $2\sigma$ by \gmtwo\ in the 
($m_{\slep_{L,R}}$, \mchi) plane for the stau-coannihilation region of the MSSM. We assume a large value of \tanb.
The orange band shows the case with $\mu= 10\tev$ and the yellow band the one with 
$\mu= 1\tev$. The solid dark blue line shows the approximate 95\%~C.L. exclusion bound from 2-lepton searches at the LHC.
\protect\subref{fig:b} Same as \protect\subref{fig:a}, for a fixed value $\mu= 5\tev$. 
The orange band shows the case with $\tanb=50$ and the yellow band the one with 
$\tanb=10$. \protect\subref{fig:c} The parameter space allowed at $2\sigma$ by \gmtwo\ in the 
($m_{\slep_R}$, \mchi) plane for the left slepton-coannihilation region of the MSSM with fixed $\tanb=50$. 
The color code is the same as in \protect\subref{fig:a}. \protect\subref{fig:d} The parameter space allowed at $2\sigma$ by \gmtwo\ in the 
($m_{\slep_R}$, \mchi) plane for the left slepton-coannihilation region of the MSSM with fixed $\mu=5\tev$.
The color code is the same as in \protect\subref{fig:b}.} 
\label{fig:slepco}
\end{figure}

In \reffig{fig:slepco}\subref{fig:a} we show the $2\sigma$ allowed parameter space in the ($m_{\slep_{L,R}}$, \mchi) plane 
for the stau-coannihilation region, in which 
the neutralino coannihilates in the early Universe with the lightest stau.
The dominant contribution to \deltagmtwomu\ is given in this case by the 
neutralino/smuon loop of Eq.~(\ref{neutsmu}), which increases linearly with $\mu$.
Thus, values of $\mu$ much larger than in the previous cases are allowed
and they actually help to satisfy the \gmtwo\ constraint.  
The bounds for two very different values of $\mu$ are shown in the plot, $\mu=1\tev$ (yellow band) and $\mu=10\tev$ (orange band), 
while $\tanb$ is kept large.

In \reffig{fig:slepco}\subref{fig:b}, the value of $\mu$ is instead fixed at an intermediate value, $\mu=5\tev$,
and we show the bounds for two different values of \tanb:  $\tanb=10$ (yellow band) and $\tanb=50$ (orange band).

We show in both panels the case with $\mtwo=5\tev$ and $m_{\slep_{L}}=m_{\slep_{R}}$, to maximize the contribution of Eq.~(\ref{neutsmu}).
Note that since Eq.~(\ref{neutsmu}) does not depend on \mtwo, the wino soft mass can be actually decoupled and 
without additional assumptions on the mechanism of SUSY breaking 3-lepton searches 
are not in principle sensitive to this region of the parameter space.
On the other hand, one can see in Figs.~\ref{fig:slepco}\subref{fig:a} and \ref{fig:slepco}\subref{fig:b} that 
the \gmtwo\ constraint bounds the value of the smuon (and selectron) masses, but the sensitivity in 2-lepton searches is still
very limited to bite significantly into the parameter space.
Note, finally, that the \gmtwo\ bounds on the left- and right- handed slepton masses presented in 
Figs.~\ref{fig:slepco}\subref{fig:a} and \ref{fig:slepco}\subref{fig:b} 
become weaker in cases where $\mu$ and \mtwo\ are both beneath $\sim 1\tev$, as 
Eq.~(\ref{charsneu}) becomes then dominant.

Rather than with the lightest stau, the neutralino can coannihilate with a light selectron, smuon, or sneutrino.
The \gmtwo\ bounds in the ($m_{\slep_R}$, \mchi) plane for the case of coannihilation with a left-handed slepton 
of the first or second generation is shown 
in Figs.~\ref{fig:slepco}\subref{fig:c} and \ref{fig:slepco}\subref{fig:d}. 
One can obtain similar plots, which well approximate the case of coannihilation with the 
right-handed selectron or smuon, by replacing $m_{\slep_R}\rightarrow m_{\slep_L}$ (Eq.~(\ref{neutsmu}) 
is symmetric under $m_{\slep_R}\leftrightarrow m_{\slep_L}$). Again we show in \reffig{fig:slepco}\subref{fig:c} 
the case of large \tanb\ for different values of $\mu$, and in 
\reffig{fig:slepco}\subref{fig:d} the case of fixed $\mu$ for different values of \tanb.
For coannihilation to occur the left-handed slepton mass is kept relatively low, $m_{\slep_L}\approx\mchi$, so that the \gmtwo\ upper bounds 
on the right-handed mass are actually weaker than in the stau-coannihilation region: $m_{\slep_R}\lesssim 2-2.5\tev$ in this case.   

The dark blue solid lines in \reffig{fig:slepco} show the approximate bound on the left-handed selectron mass from the LHC 2-lepton searches for 
$\tilde{e}_L \tilde{e}_L$ pair production. In the cases shown in Figs.~\ref{fig:slepco}\subref{fig:c} and \ref{fig:slepco}\subref{fig:d} 
the reader should take note of a couple of caveats: the first is that the bound does not properly apply when 
a left-handed selectron is degenerate with the neutralino because the spectrum is compressed; the second is that 
the limit for right-handed slepton pair production is actually weaker than the one shown here by $\sim50\%$,
because the cross section for $\tilde{e}_R \tilde{e}_R$ production is suppressed with respect to left-handed production.
The proper and complete treatment of the LHC limits for the cases with coannihilation will be presented in \refsec{sec:results}.\medskip

\noindent \textbf{Nearly pure higgsinos and winos}.
We finally discuss the parameter space corresponding to an apparent underabundance of DM in the Universe, $\abundchi\lesssim 0.12$.
As is well known, this situation is typical when the LSP is an almost pure higgsino with $\mchi\ll 1\tev$\cite{Roszkowski:1991ng,Profumo:2004at}
or an almost pure wino with $\mchi\ll 2.8\tev$\cite{Hryczuk:2010zi}. These solutions are generally thought to be proper of 
scenarios with two-component DM, or of cases where the neutralino represents the entirety of the DM 
and the correct abundance is fixed by invoking some additional mechanism, e.g., freeze-in\cite{Hall:2009bx}.   

\begin{figure}[t]
\centering
\subfloat[]{%
\label{fig:a}%
\includegraphics[width=0.47\textwidth]{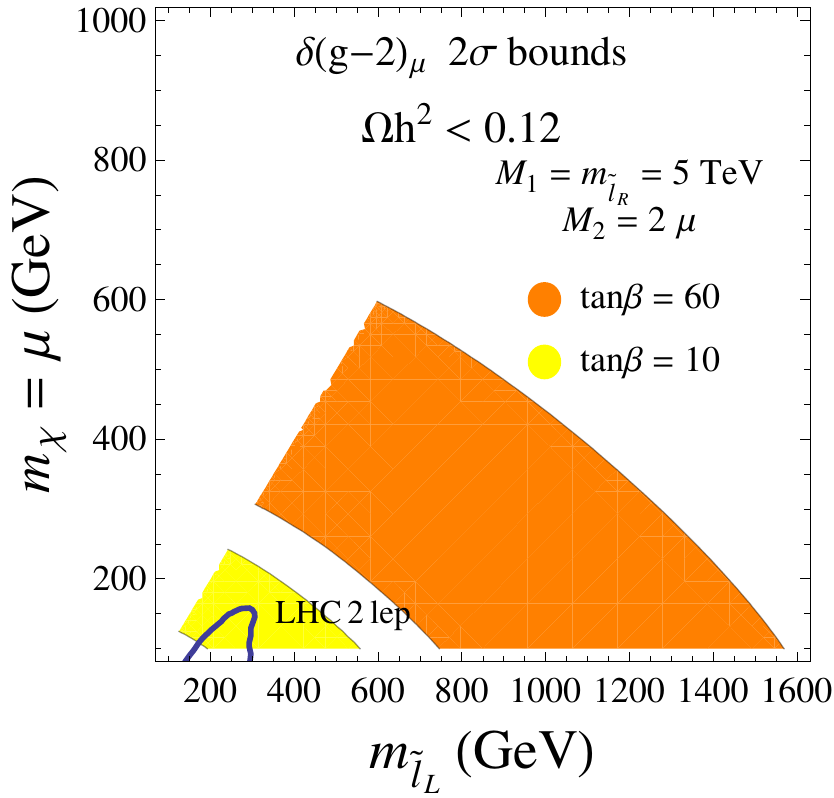}
}%
\hspace{0.02\textwidth}
\subfloat[]{%
\label{fig:b}%
\includegraphics[width=0.47\textwidth]{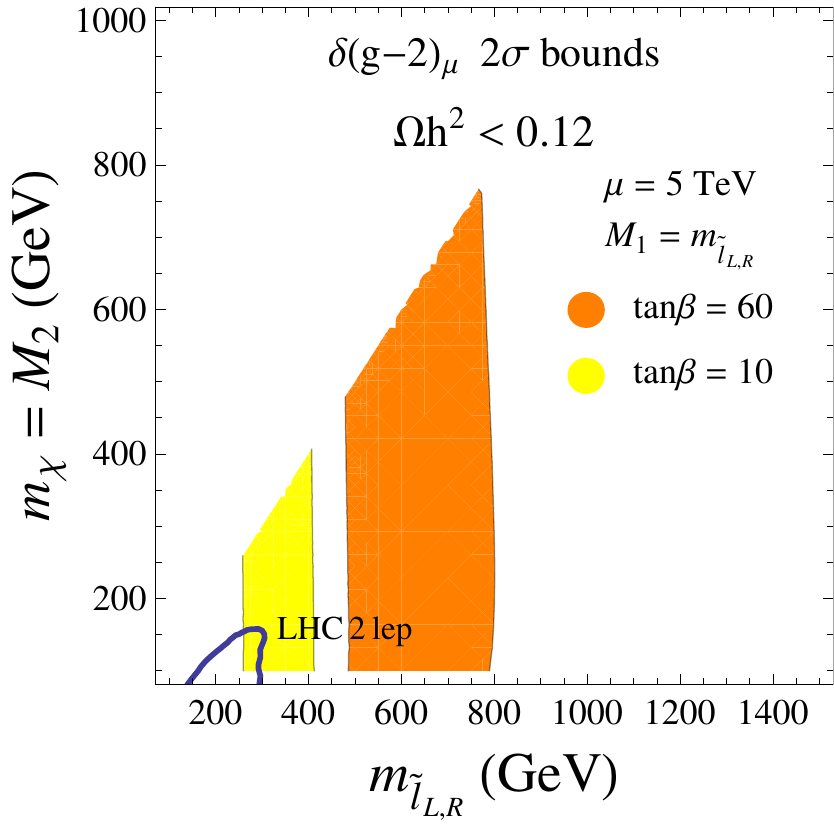}
}%
\caption{\footnotesize The parameter space allowed by \gmtwo\
at $2\sigma$ in the ($m_{\slep_L}$, \mchi) plane for some cases with $\abundchi\lesssim 0.12$. 
\protect\subref{fig:a} Case of a higgsino-like neutralino with $\mchi\approx\mu<1\tev$. 
\protect\subref{fig:b} Case of a wino-like neutralino with $\mchi\approx \mtwo\ll 3\tev$.
In both plots the orange band shows the case with $\tanb=60$ and the yellow band the one with $\tanb=10$. The solid dark blue line approximately shows the
95\%~C.L. exclusion bound from 2-lepton searches at the LHC.}  
\label{fig:pures}
\end{figure}

We show in \reffig{fig:pures}\subref{fig:a} the $2\sigma$ allowed parameter space in the ($m_{\slep_L}$, \mchi)
plane for a higgsino LSP. The plot is generated under the assumptions that \mtwo\ is greater than $\mu$, but it cannot be \textit{orders of magnitude} greater, otherwise the dominant contribution to \deltagmtwomu, Eq.~(\ref{charsneu}), becomes drastically suppressed, 
as was the case for the mixed bino-higgsino scenario. 
On the other hand, Eq.~(\ref{charsneu}) is insensitive to the values of $m_{\slep_R}$ and \mone, which can be decoupled.

In \reffig{fig:pures}\subref{fig:b} we show the allowed parameter space for a wino-like neutralino with relatively large $\mu$.
Besides Eq.~(\ref{charsneu}), substantial contribution to \deltagmtwomu\ comes in this case from 
Eq.~(\ref{neutsmu}), so that \deltagmtwomu\ can be enhanced for large $\mu$ when the slepton and bino masses are not much above \mtwo. 
For smaller values of $\mu$ the behaviour becomes similar to that of the pure higgsino and mixed bino-higgsino cases.  

As was the case for the bino/higgsino admixtures described above, almost pure higgsinos
and winos are extremely difficult to test at the LHC in 3-lepton final state searches,
because of the strong degeneracy between \charone\ and $\chi$.  
On the other hand, it is also clear from \reffig{fig:pures} that 
2-lepton searches can begin to test the parameter space in these cases but, especially for larger values of \tanb,
it will be very hard to reach enough sensitivity to probe the full allowed parameter space.

To summarize, this discussion has proven that, when one requires consistency with the \gmtwo\ constraint and the relic density,
important regions of the MSSM parameter space are within the reach of the LHC. 
However, in the MSSM there is too much freedom and one cannot entirely test the hypothesis of a SUSY origin 
to the \gmtwo\ anomaly at the LHC, 
especially in cases where degeneracy among the main parameters becomes important.
In what follows, we will focus on theoretically well-motivated scenarios where the presence of additional 
symmetries at the GUT scale removes much of the freedom of the parameter space.
We limit ourselves to the most common case of a bino-like neutralino, which can saturate the relic density and is testable at the LHC,
but we will comment on the cases that can give rise to neutralinos of a different composition.


\section{GUT-defined models and experimental constraints\label{sec:scenarios}}

Many models of gravity mediation assume for simplicity universal GUT-scale conditions for the gaugino and scalar soft masses.
As was mentioned in \refsec{sec:intro}, however, under those assumptions it has become no longer possible to find 
parameter space consistent with the \gmtwo\ constraint after the Higgs discovery and null searches for squarks and gluinos at LHC run I. 
As a matter of fact, in scans of the CMSSM and the NUHM the \gmtwo\ constraint is in some cases neglected\cite{Fowlie:2012im,Roszkowski:2014wqa}
when looking for the regions of the parameter space favored by the Higgs measurement or the LHC. 
The issue is somewhat controversial and judgement is in general postponed to after an eventual confirmation 
by more precise upcoming experiments.

We follow here a different approach and consider GUT-defined models that actually 
\textit{do} satisfy the present constraints
for \gmtwo. To this end, we relax the assumption of gaugino universality. 
We will show in the remainder of this paper that
the LHC bounds on the EW sector of these GUT-defined SUSY models are quite strong and
that the 14\tev\ run will be able to probe the parameter space of these models virtually in its entirety.
 
\begin{table}[t]
\begin{center}
\begin{tabular}{|c|c|c|}
\hline
\textbf{Model 1} & \textbf{CMSSM-like $\boldsymbol{M_3}$ floating} & \\
\hline
\footnotesize Parameter & \footnotesize Description & \footnotesize Range \\
\hline
\mzero\ & \footnotesize Universal scalar mass & $100,\,4000$ \\
\hline
\mhalf & \footnotesize Bino/wino soft mass & $100,\,4000$ \\
\hline
$M_3$ & \footnotesize Gluino soft mass & $700,\,10000$ \\
\hline
\azero\ & \footnotesize Universal trilinear coupling & $-\,8000,\,8000$ \\
\hline
\tanb\ & \footnotesize Ratio of the Higgs vevs & $2,\,62$ \\
\hline
\signmu\ & \footnotesize Sign of the Higgs/higgsino mass parameter & $+\,1$ \\
\hline
\hline
\textbf{Model 2} & \textbf{Non-universal gaugino masses} & \\
\hline
\mone & \footnotesize Bino soft mass & $-\,4000,\,4000$ \\
\hline
\mtwo & \footnotesize Wino soft mass & $-\,4000,\,4000$ \\
\hline
\mzero, $M_3$, \azero, \tanb, \signmu\ & \footnotesize Same as Model~1 & \footnotesize Same as Model~1 \\
\hline
\hline
\textbf{Model 3} & \textbf{$\boldsymbol{SO(10)}$-like sfermions} & \\
\hline
$m_{16}$ & \footnotesize Universal scalar mass \textbf{16} repr. & $100,\,4000$ \\
\hline
$m_{10}^2$ & \footnotesize Universal scalar mass \textbf{10} repr. & $-\,10000^2,\,10000^2$ \\
\hline
$3\,M_D^2$ & \footnotesize $D$-term extra $U(1)$ & $0,\,m_{16}^2-(100\gev)^2$ \\
\hline
\mhalf\ & \footnotesize Bino/wino soft mass  & $100,\,2000$ \\
\hline
$M_3$ & \footnotesize Gluino soft mass & $800,\,5000$ \\
\hline
\azero, \tanb, \signmu\ & \footnotesize Same as Model~1 & \footnotesize Same as Model~1 \\
\hline
\hline
\textbf{Model 4} & \textbf{$\boldsymbol{SU(5)}$-like sfermions} & \\
\hline
$m_{10}$ & \footnotesize Universal scalar mass \textbf{10} repr. & $100,\,4000$ \\
\hline
$m_{5}$ & \footnotesize Universal scalar mass $\boldsymbol{\bar{5}}$ repr. & $100,\,2000$ \\
\hline
\mhdsq\ & \footnotesize Down Higgs doublet soft mass & $-\,10000^2,\,10000^2$ \\
\hline
\mhusq\ & \footnotesize Up Higgs doublet soft mass & $-\,10000^2,\,10000^2$ \\
\hline
\mhalf, $M_3$, \azero, \tanb\, \signmu & \footnotesize Same as Model~3  & \footnotesize Same as Model~3 \\
\hline
\textbf{Model 4-zoom} & \textbf{$\boldsymbol{SU(5)}$ $\boldsymbol{\mu, \ma}$ parameterization} & \\
\hline
$\mu$ & \footnotesize EW-scale higgsino mass parameter & $10,\,2000$ \\
\hline
\ma\ & \footnotesize Pseudoscalar pole mass & $100,\,4000$ \\
\hline
$M_3$ & \footnotesize Gluino soft mass & $500,\,2000$ \\
\hline
$m_{10}$, $m_{5}$, \mhalf, \azero, \tanb\ & \footnotesize Same as Model~4  & \footnotesize Same as Model~4 \\
\hline
\end{tabular}
\caption{Parameters of the models analyzed in this work. All soft SUSY-breaking masses are defined at the GUT scale. 
Dimensionful quantities are given in GeV and $\gev^2$.}
\label{tab:models}
\end{center}
\end{table}%

The models we analyze are summarized in \reftable{tab:models} and discussed below.
The scans are performed with the package BayesFITS\cite{Fowlie:2012im,Fowlie:2013oua,Roszkowski:2014wqa,Roszkowski:2014iqa} 
which interfaces several publicly available tools to 
direct the scanning procedure and calculate physical observables. 
The sampling is performed by \texttt{MultiNest}\cite{Feroz:2008xx} with 4,000 or 6,000 live points. 
We use $\tt SoftSusy \,v.3.5.2$\cite{Allanach:2001kg} to calculate the mass spectrum and \texttt{SUSY-HIT}\cite{Djouadi:2006bz} for the decay branching ratios. 
Higher-order corrections to the Higgs mass are calculated with 
$\tt FeynHiggs\ v.2.10.2$
\cite{Heinemeyer:1998yj,Heinemeyer:1998np,Degrassi:2002fi,Frank:2006yh,Hahn:2013ria}.
$\tt FeynHiggs$ is interfaced with $\tt HiggsSignals\,v1.3.1$\cite{Bechtle:2013xfa} and $\tt HiggsBounds\,v4.2.0$\cite{Bechtle:2008jh,Bechtle:2011sb,Bechtle:2013wla} 
to evaluate the constraints on the Higgs sector. $\tt SuperISO\ v.3.4$\cite{Mahmoudi:2008tp} 
is used to calculate \deltagmtwomu\ and flavor observables \brbxsgamma, \brbsmumu, and \brbutaunu.
$M_W$, \sinsqeff, \delmbs\ are calculated with $\tt FeynHiggs$. 
Dark matter observables, \abundchi\ and the spin-independent DM-proton cross section, \sigsip,
are computed with $\tt micrOMEGAs\ v.4.1.5$\cite{Belanger:2013oya}. 

$\tt SuperISO\ v.3.4$ performs the calculation of \deltagmtwomu\ including
the leading-log QED corrections from 2-loop evaluations\cite{Degrassi:1998es}, 
photonic Barr-Zee diagrams with physical Higgs\cite{Chang:2000ii,Chen:2001kn,Arhrib:2001xx} 
and bosonic EW 2-loop contributions\cite{Heinemeyer:2004yq}. 

\begin{table}[t]
\begin{center}
\begin{tabular}{|c|c|c|c|c|}
\hline
Constraint & Mean & Exp. Error & Th. Error & Ref. \\
\hline
Higgs sector & See text. & See text. & See text. & \cite{Bechtle:2013xfa,Bechtle:2008jh,Bechtle:2011sb,Bechtle:2013wla} \\
\hline
LUX & See\cite{Kowalska:2014hza,Roszkowski:2014wqa}. & See\cite{Kowalska:2014hza,Roszkowski:2014wqa}. & See\cite{Kowalska:2014hza,Roszkowski:2014wqa}. & \cite{Akerib:2013tjd}\\
\hline
\abunchi\ & 0.1199 & 0.0027 & 10\% & \cite{Ade:2013zuv}\\
\hline
$\deltagmtwomu\times 10^{10}$ & 28.7 & 8.0 & 3.0 & \cite{Bennett:2006fi,Miller:2007kk}\\
\hline
\sinsqeff\ & 0.23155 & 0.00015 & 0.00015 &  \cite{Beringer:1900zz} \\
\hline
$\brbxsgamma\times 10^4$ & 3.43 & 0.22 & 0.21 & \cite{bsgamma} \\
\hline
$\brbutaunu \times 10^4$ & 0.72 & 0.27 & 0.38 & \cite{Adachi:2012mm} \\
\hline
\delmbs\ & 17.719~ps$^{-1}$ & 0.043~ps$^{-1}$ & 2.400~ps$^{-1}$ & \cite{Beringer:1900zz} \\
\hline
$M_W$ & $80.385\gev$ & $0.015\gev$ & $0.015\gev$ & \cite{Beringer:1900zz} \\
\hline
$\brbsmumu \times 10^9$ & 2.9 & $0.7$ & 10\% & \cite{Aaij:2013aka,Chatrchyan:2013bka} \\
\hline
$\Gamma(Z\rightarrow\chi\chi)$ & $\leq 1.7\mev$ & $0.3$ & -- & \cite{Agashe:2014kda} \\
\hline
\end{tabular}
\caption{The experimental constraints applied in this study.}
\label{tab:exp_constraints}
\end{center}
\end{table}%

The scans are subject to a set of constraints, applied through a global likelihood function $\mathcal{L}$.
The list of constraints, central values, theoretical and experimental uncertainties are 
presented in Table~\ref{tab:exp_constraints}. We assume Gaussian distributions for the constraints, 
with the exception of those on the Higgs sector, which are imposed through $\tt HiggsSignals$ and $\tt HiggsBounds$, and the constraints
on \sigsip\ from LUX\cite{Akerib:2013tjd}. 
The LUX constraint, which slightly improved on the limit from XENON100\cite{Aprile:2012nq}, 
is included in the likelihood function following the procedure detailed in\cite{Cheung:2012xb,Fowlie:2013oua,Kowalska:2014hza}. 
Additionally, we impose 95\%~C.L. lower bounds
from direct searches at LEP\cite{Agashe:2014kda}, smeared with 5\% theoretical errors. 
The limits are given in Eq.~(2) of Ref.\cite{Fowlie:2013oua}, 
with the exception of the limit on the neutralino mass 
that has been replaced here by the LEP limit on the invisible $Z$ width, $\Gamma(Z\rightarrow\chi\chi)$\cite{Agashe:2014kda}.

To define the $2\sigma$ allowed regions, we adopt for \deltagmtwomu\ the central value of Eq.~(\ref{measure1}). We estimate
the theoretical uncertainty very conservatively, $\sigma_{\textrm{th}}^{(a_{\mu})}=3.0\times 10^{-10}$, to 
bundle together the uncertainties that arise from neglecting hadronic 2-loop corrections\cite{Fargnoli:2013zda,Fargnoli:2013zia} 
in the SUSY calculation and the SM uncertainties that give rise to different estimates, like Eq.~(\ref{measure2}).\smallskip  

The first model we consider, \textbf{Model~1} hereafter, is a simple modification to the CMSSM first introduced 
in Ref.\cite{Akula:2013ioa}.\footnote{The authors of Ref.\cite{Akula:2013ioa} call these scenarios $\tilde{g}$SUGRA, as the radiative breaking of EW symmetry is driven by the gluino, $\tilde{g}$.}
In addition to the usual parameters, the gluino soft mass $M_3$ is allowed to float at the GUT scale, as shown in \reftable{tab:models}. 
This is the minimal implementation of non-universality that allows one to simultaneously respect the bounds on 
the color sector from the LHC and the Higgs mass, and those on the EW sector from \gmtwo.
As was shown in Ref.\cite{Akula:2013ioa}, this simple condition can be easily obtained within several GUT symmetries. 

The distributions of the input parameters of Model 1 after applying the constraints of Table~3 are not particularly 
illuminating for the purposes of this paper and we refrain from showing them here.
As expected, at the GUT scale $M_3$ assumes large values, $M_3\simeq 1-5\tev$, constrained by
the fact that it must drive the physical stop masses to the multi-TeV regime to comfortably fit  
the Higgs mass and rates. One finds the following approximate relations between the physical masses and the GUT-scale value of $M_3$: 
$\mstopone\approx 1.5\,M_3$, $\mglu\approx 2\,M_3$.
Conversely, the common scalar mass remains small, $\mzero\lesssim 450\gev$, due to the fact that the physical smuon mass must be small enough to 
be consistent with \deltagmtwomu.
 
\begin{figure}[t]
\centering
\subfloat[]{%
\label{fig:a}%
\includegraphics[width=0.50\textwidth]{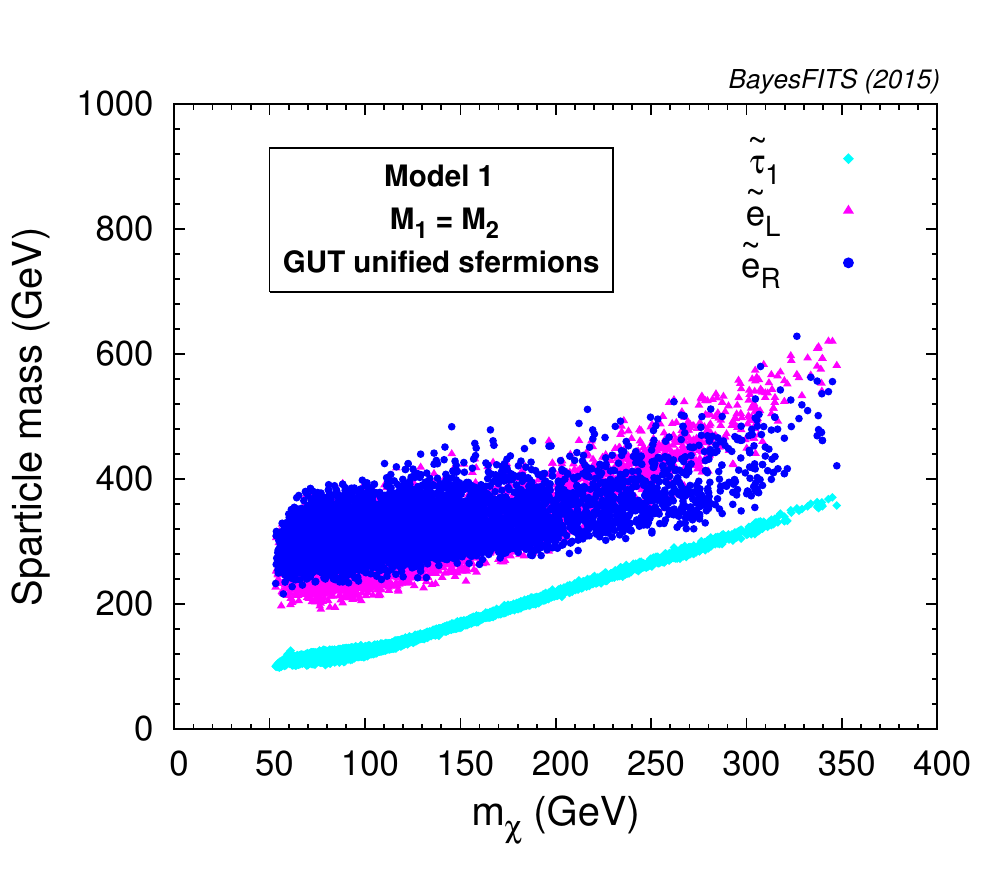}
}%
\\
\subfloat[]{%
\label{fig:b}%
\includegraphics[width=0.47\textwidth]{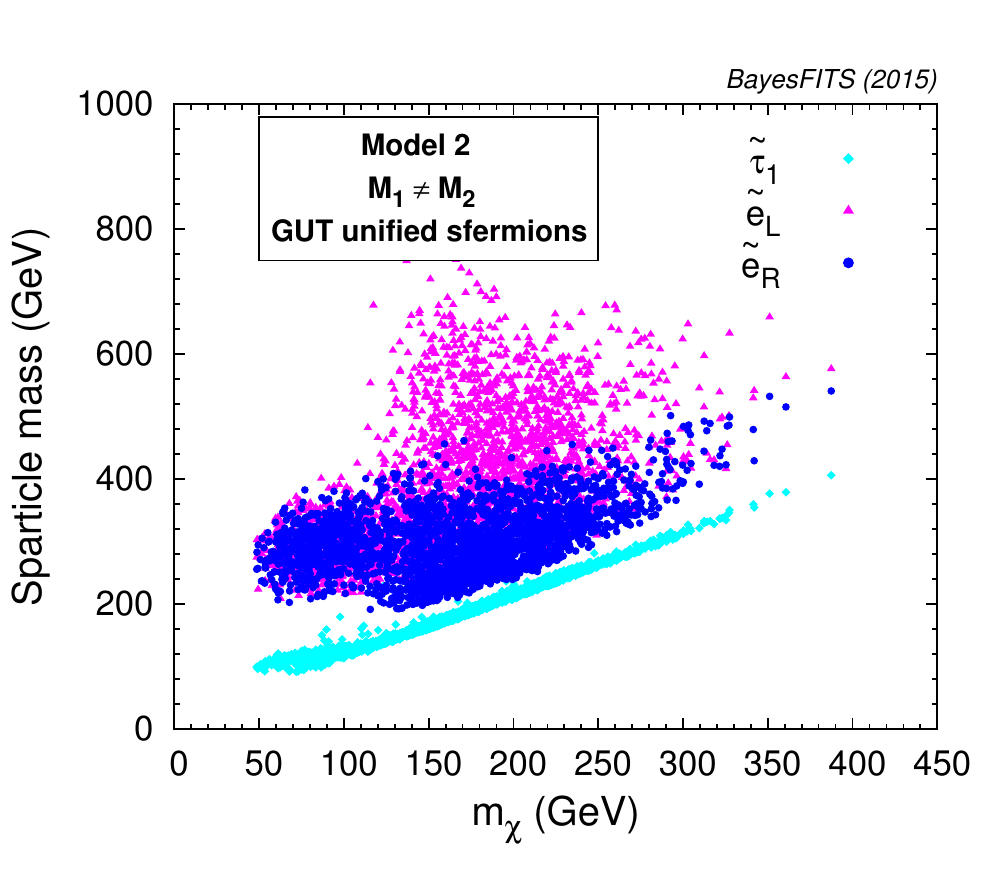}
}%
\hspace{0.02\textwidth}
\subfloat[]{%
\label{fig:c}%
\includegraphics[width=0.47\textwidth]{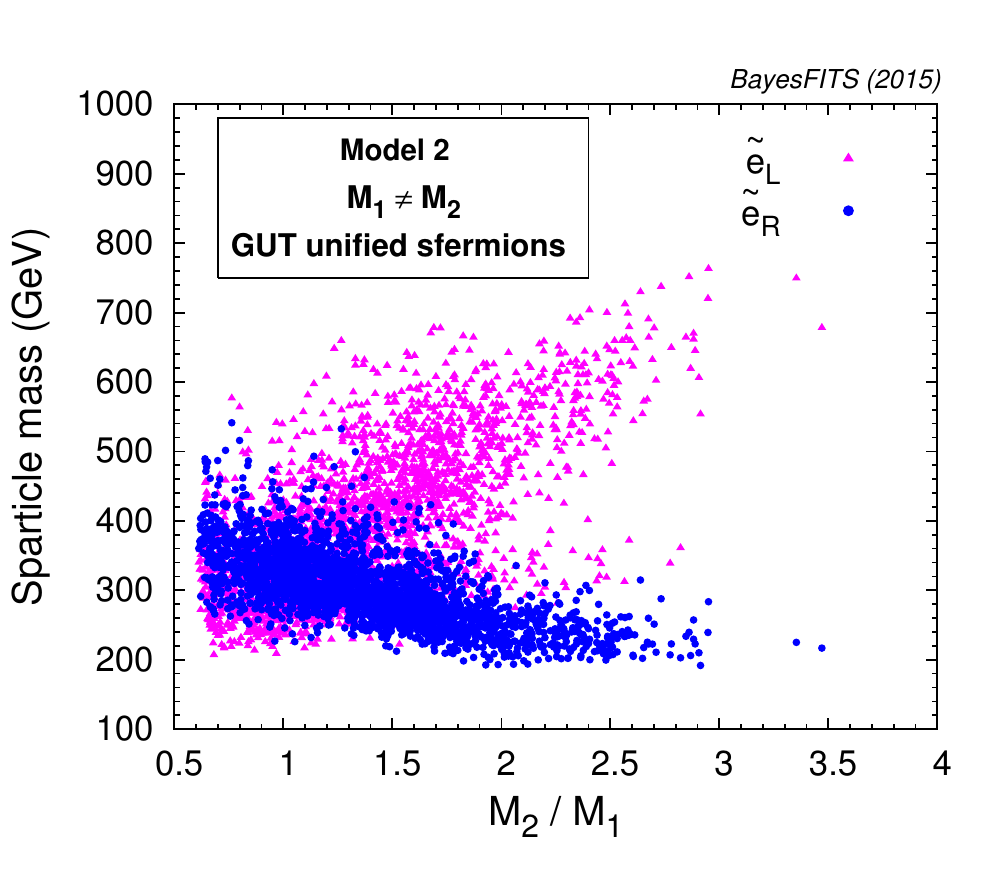}
}%
\caption{\footnotesize \protect\subref{fig:a} The distribution of the physical left-handed selectron mass (magenta triangles), 
right-handed selectron mass (blue circles), 
and lightest stau mass (cyan diamonds) versus the neutralino mass for the points of Model~1 that satisfy the constraints of \reftable{tab:exp_constraints}
at the $2\sigma$ level. \protect\subref{fig:b} Same as \protect\subref{fig:a} for Model~2.
\protect\subref{fig:c} The distribution of the physical left-handed selectron mass (magenta triangles) and 
right-handed selectron mass (blue circles) versus the ratio of the GUT scale value of the wino soft mass to the bino soft mass, $\mtwo/\mone$, for Model~2.}  
\label{fig:relic}
\end{figure}

As is often the case in global SUSY analyses, the relic density provides the constraint with the strongest impact on the EW sector.
In \reffig{fig:relic}\subref{fig:a} we plot with magenta triangles the distribution of the 
physical left-handed selectron mass, $m_{\tilde{e}_L}$, versus the neutralino mass, \mchi, 
for the points of Model~1 satisfying the constraints of \reftable{tab:exp_constraints} at the $2\sigma$ level.  
The right-handed selectron mass distribution, $m_{\tilde{e}_R}$ is shown with blue circles, 
and the lightest stau mass distribution, $m_{\stau_1}$, with cyan diamonds.
One should remember that in gravity-mediated models $m_{\tilde{\mu}_L}\approx m_{\tilde{e}_L}$, given the 
very small value of the Yukawa couplings, and the same is true for the right-handed sleptons.   

The neutralino is strongly bino-dominated and the correct relic abundance is obtained, 
for $100\gev\lesssim\mchi\lesssim 350\gev$, through coannihilation with the lightest stau.
For a lighter \mchi\ it is obtained through bulk-like annihilation\cite{Griest:1988ma} to taus via $t$-channel exchange of the moderately light stau.       
One can see that coannihilation rapidly loses efficiency as \mchi\ increases,
so that no solutions are found for $\mchi\gsim 350\gev$.
 
As was explained in \refsec{sec:g2andrelic} the dominant contribution to \deltagmtwomu\ is given by the neutralino/smuon loop of Eq.~(\ref{neutsmu})
so that the parameter $\mu$ adopts fairly large values, $\mu\gsim 2-5\tev$. 
Unlike in general low-scale MSSM scenarios like the one described in \refsec{sec:g2andrelic},
in Model~1 \tanb\ can only assume moderate values, $\tanb\simeq5-25$, as for larger \tanb\ the stau masses run 
to unphysical low-scale values giving $m_{\stau_1}<\mchi$.\smallskip  

The GUT-scale universality condition $\mone=\mtwo\equiv\mhalf$ is relaxed in \textbf{Model~2},
whose parameters are shown in \reftable{tab:models}.
We show in \reffig{fig:relic}\subref{fig:b} the distribution of the physical left-handed selectron mass, right-handed selectron mass, 
and stau mass versus the neutralino mass for the $2\sigma$ allowed points.
As can be inferred by the stau mass distribution, the mechanism of interest for the relic density is again stau-coannihilation,
like in Model~1. The main difference with the previous case is that, in addition to a broader range of $m_{\charone}$ values, there is a 
broader distribution for the slepton masses of the first two generations.

This can be understood by looking at \reffig{fig:relic}\subref{fig:c}, where we show with magenta triangles the distribution of the left-handed
selectron mass versus the GUT-scale ratio $\mtwo/\mone$, and with blue circles the corresponding distribution 
for the right-handed selectron mass.
Note that we can only find solutions in the $2\sigma$ region of the constraints of 
\reftable{tab:exp_constraints} when $\mtwo/\mone>0$. 
Model~1 is the subset of Model~2 represented by the points at $\mtwo/\mone=1$.
One can see that the splitting between the left- and right-handed selecton masses increases with increasing $\mtwo/\mone$,
as larger $\mtwo$ values can drive the left-handed mass to larger values at the low scale through the RGEs. 
Moreover, for ratios larger than $\mtwo/\mone\simeq 3.5$ 
the $\stau_L$ soft mass becomes too large after running to the low-scale to allow for efficient coannihilation with the neutralino,
even in the presence of large stau mixing. At the same time it becomes difficult to accommodate an increasing 
$m_{\tilde{\mu}_L}$ within the constraints from \gmtwo\ so that no additional points with larger $\mtwo/\mone$ 
can be found in the $2\sigma$ region.

Conversely, for $\mtwo/\mone\ll 1$ the LSP becomes wino-like and \abundchi\ drops down to very small values.
As was explained at the end of \refsec{sec:g2andrelic}, wino-like charginos and neutralinos are highly degenerate,
which makes their detection very challenging at the LHC.
Additionally light wino-like neutralinos require an extra dark matter component or production mechanisms beyond those in the MSSM to satisfy \abundchi.
For these reasons, we limit ourselves here to the analysis of the parameter space yielding predominantly a 
bino-like LSP.\smallskip 

\textbf{Model~3} introduces a small difference between the right- and left-handed soft sfermion masses at the GUT-scale.
The prototype we have in mind is a supergravity-based, GUT-scale model characterized by $SO(10)$ boundary conditions\cite{Georgi:1974my,Fritzsch:1974nn},
where we assume a small positive $D$-term contribution, $M_D^2$, from the extra (broken) $U(1)$\cite{Kawamura:1994ys,Kolda:1995iw}. 

As usual, the GUT-defined soft squark masses ($m_Q^2$, $m_U^2$, $m_D^2$), slepton masses ($m_L^2$, $m_E^2$), and Higgs doublets masses
can be parametrized in terms of the universal scalar mass in the fermionic \textbf{16} representation, $m_{16}$, 
the universal scalar mass in the bosonic \textbf{10} representation, $m_{10}$, and the $D$-term, $M_D^2$, so that   
\bea 
m_{Q}^2=m_{U}^2=m_{E}^2&\equiv & m_{16}^2+M_D^2\nonumber\\
m_{D}^2=m_{L}^2 & \equiv & m_{16}^2-3 M_D^2\nonumber\\
m_{H_{u,d}}^2&\equiv & m_{10}^2\mp 2 M_D^2\,.\label{so10cond}
\eea
 
We scan $m_{16}$, $m_{10}$, and $M_D^2$ in the ranges given in \reftable{tab:models}
and we neglect here the effects of including the right-handed sneutrino masses, as they have only slight impact on the low-energy spectrum\cite{Baer:2000gf}.
Moreover, we scan \tanb\ in the range $2-62$ to increase the number of solutions, thus 
ignoring the requirement of successful $SO(10)$ unification of the Yukawa couplings.
A study of the impact of \gmtwo\ in $SO(10)$ models including the Yukawa unification 
constraint can be found, e.g., in  Ref.\cite{Badziak:2011wm}. 

As was the case in Model~1, we assume GUT-scale unification 
of \mone\ and \mtwo\ and we leave $M_{3}$ free to float, so to fit the Higgs mass without affecting the parameters that enter \gmtwo, and so
that at the EW scale a large, gluino-driven, sbottom mass can easily evade the bounds from direct searches at the LHC.

Obviously, Model~1 is a subset of Model~3, so that the parameter space shown in 
\reffig{fig:relic}\subref{fig:a} is common to both models. On the other hand, the extra 
freedom that comes in Model~3 from the right/left splitting at the GUT scale opens up an additional region of the
low-scale parameter space, where the relic density is satisfied thanks to coannihilation of a bino-like neutralino
and an almost degenerate right-handed slepton of the first or second generation.   

\begin{figure}[t]
\centering
\subfloat[]{%
\label{fig:a}%
\includegraphics[width=0.47\textwidth]{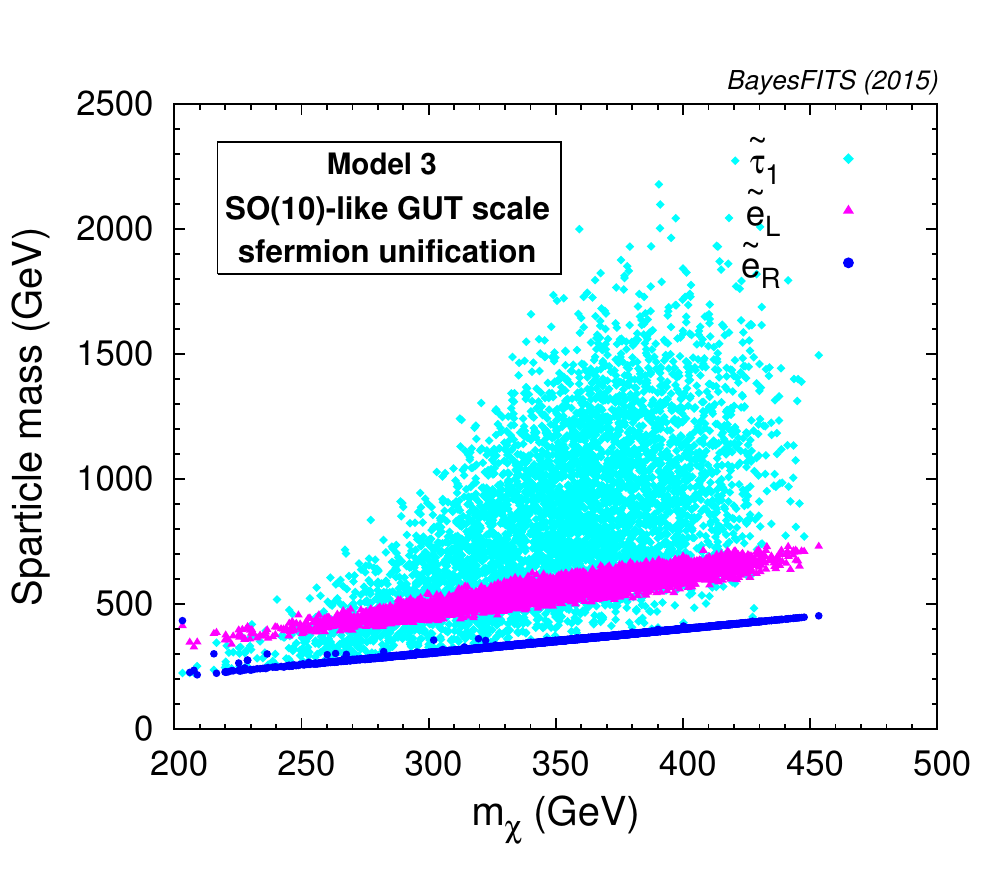}
}%
\hspace{0.02\textwidth}
\subfloat[]{%
\label{fig:b}%
\includegraphics[width=0.47\textwidth]{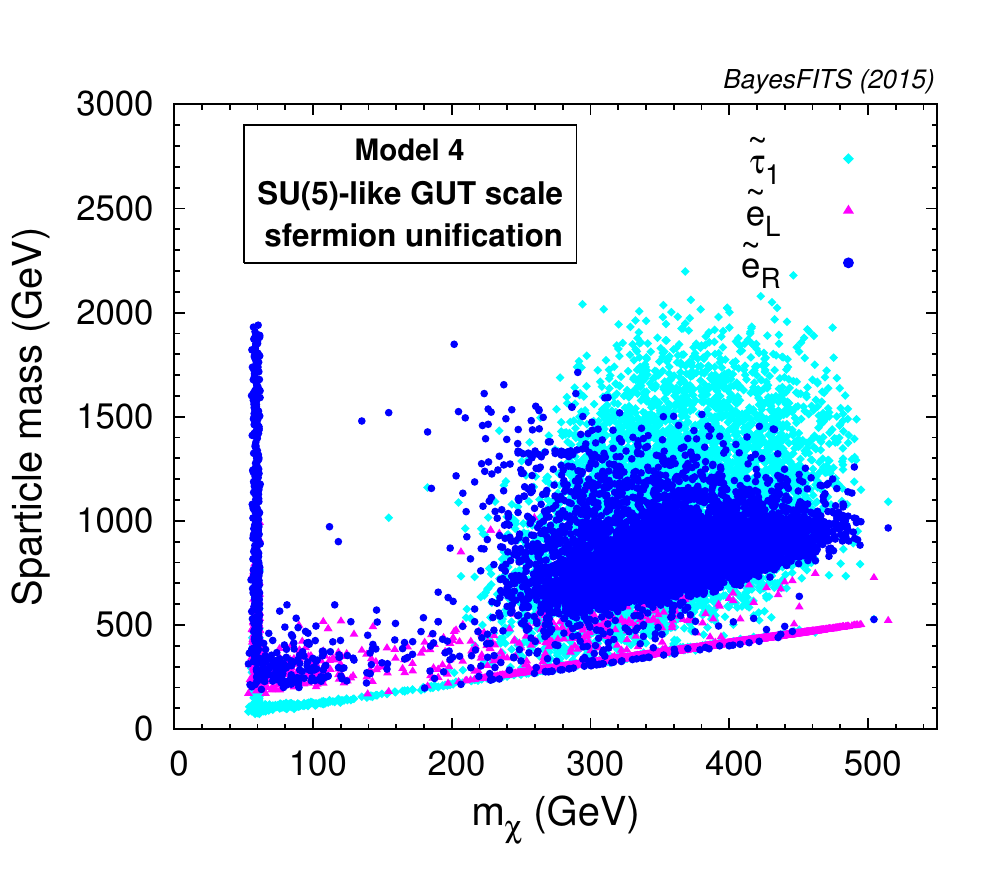}
}%
\caption{\footnotesize \protect\subref{fig:a} The distribution of the physical left-handed selectron mass (magenta triangles), 
right-handed selectron mass (blue circles), and lightest stau mass (cyan diamonds) versus the neutralino mass for the points of Model~3 that 
are \textit{not} in common with Model~1. All points satisfy the constraints of \reftable{tab:exp_constraints}
at the $2\sigma$ level. \protect\subref{fig:b} The distribution of the slepton physical masses in Model~4.}  
\label{fig:sosu}
\end{figure}

We show in \reffig{fig:sosu}\subref{fig:a} the physical $m_{\tilde{e}_L}$, $m_{\tilde{e}_R}$, and $m_{\stau_1}$ distributions 
versus the neutralino mass for the right slepton-coannihilation region.
As was the case in Model~1, an upper bound on \mchi\ can be derived, $\mchi\lesssim 450\gev$, beyond
which coannihilation becomes no longer efficient and \abundchi\ starts to rise.
The main contribution to \deltagmtwomu\ come from Eq.~(\ref{neutsmu}) and in this region $\mu\simeq4-10\tev$.
Figure~\ref{fig:sosu}\subref{fig:a} also shows that in this region the left-handed sleptons and, for many points,
even the lightest staus, are not much heavier than the neutralino (and the lightest chargino, which is wino-like with 
$m_{\charone}\approx 2\,\mchi$). As we shall see in \refsec{sec:results}, this has important consequences when it comes to the 
LHC signatures.\smallskip

\textbf{Model 4} generalizes Model~3 but extends the allowed GUT-scale ranges for the right- and left-handed sleptons.
The prototype we have in mind is a supergravity model with minimal $SU(5)$ boundary conditions at the GUT scale\cite{Georgi:1974sy}.
The GUT-defined soft squark and slepton  masses are parametrized in terms of the 
common scalar soft mass for the fields belonging to the $\mathbf{\boldsymbol{\bar{5}}}$ representation, $m_{5}$, and the common scalar
mass in the \textbf{10}, $m_{10}$. The Higgs doublets' soft masses are free, so that one has overall,  
\bea 
m_{Q}^2=m_{U}^2=m_{E}^2&\equiv & m_{10}^2\nonumber\\
m_{D}^2=m_{L}^2 & \equiv & m_{5}^2\,,\label{su5cond}
\eea
and \mhusq, \mhdsq\ are independent free parameters.
Again, we assume that at the GUT scale $\mone=\mtwo\equiv \mhalf$ and $M_3$ is free-floating. 
The parameter ranges for $m_{5}$, $m_{10}$, \mhusq, \mhdsq\ can be found in
\reftable{tab:models}.

Additionally, we also perform a scan of the parameter space of Model~4 after trading 
the GUT-scale inputs \mhusq\ and \mhdsq\ for the low-scale defined $\mu$ parameter and the
pseudoscalar pole mass, $m_A$. We restrict the range of $\mu$ to $10-2000\gev$, and at the same time we restrict 
the range of $M_3$ to $500-2000\gev$ (this scan is called Model~4-zoom in \reftable{tab:models}). 
We do this to counterbalance the tendency of our scans to find solutions characterized 
by very large values of $\mu$ and $M_3$. 
In the spirit of phenomenology we will not consider issues of EW naturalness in this study.
However, it is interesting to see if solutions more natural than the ones generated in Models~1-3
are possible. The scan of Model~4-zoom is designed to expose the region 
characterized by a Barbieri-Giudice measure\cite{Ellis:1986yg,Barbieri:1987fn} roughly less than 1000 
(see, e.g., Ref.\cite{Kowalska:2014hza} for a discussion).

Models~1 and 3 are sub-cases of Model~4, so that they share common regions of the parameter space. 
However, Model~4 introduces more freedom to the parameter space, 
with the consequence that there are regions for which the relic density is satisfied thanks to coannihilation of a
left-handed slepton of the first or second generation and the neutralino.
This is shown on the right of \reffig{fig:sosu}\subref{fig:b}, where we plot the physical slepton 
mass distributions for Models~4 and 4-zoom.
Additionally, one can see on the left of \reffig{fig:sosu}\subref{fig:b}, for $\mchi\simeq 60\gev$,
some solutions belonging to the $h$-resonance region, which are efficiently explored by scanning the parameters
in the limited ranges of Model 4-zoom.

We want to point out here that the $SU(5)$ boundary conditions (\ref{su5cond}) of Model~4 are not the ones exclusively giving the 
described low energy phenomenology. Because the GUT-scale value of $M_3$ is allowed to be large and drive the squarks 
to decoupled values, the same solutions shown in \reffig{fig:sosu}\subref{fig:b}
also apply to cases with different symmetries at the GUT scale, as would be the case, e.g.,
of Pati-Salam\cite{Pati:1974yy} boundary conditions. 
  
\section{Constraints and projections for the LHC\label{sec:results}}

In this section we confront the GUT-defined scenarios discussed in \refsec{sec:scenarios} 
with the bounds from direct SUSY searches at the LHC. 
In \refsec{sec:searches} we briefly describe the numerical methodology we employ to derive the bounds 
from the 8\tev\ run and we show our projected sensitivities for the 14\tev\ run in SMS scenarios. 
In the following subsections we move on to discuss the impact of the implemented searches on 
the allowed parameter space of the considered models.

\subsection{Numerical implementation of LHC searches\label{sec:searches}}

We numerically reproduce three LHC searches designed to explore the EW sector of the MSSM: 
the searches for EW chargino and neutralino production with 3 leptons in the final state by ATLAS and 
CMS\cite{Aad:2014nua,Khachatryan:2014qwa} (collectively called ``3-lepton" hereafter), and a search for direct slepton pair 
production, sneutrino pair production, and slepton/sneutrino production with two opposite-sign leptons in the final state, 
by ATLAS\cite{Aad:2014vma} (dubbed as ``2-lepton" hereafter). 
In 3-lepton analyses, CMS and ATLAS both reported at the end of the 8\tev\ run 
small excesses in the observed events in different signal regions. Thus, each collaboration 
presented exclusion bounds slightly weaker than the expected ones, albeit not in the same signal regions. 
To take advantage of the stronger limits from each collaboration we adopt here a ``best of" strategy 
when we impose the limits from 3 lepton searches on the parameter space of our models.

To recast the 3-lepton and the 2-lepton ATLAS searches, we employ the publicly 
available code~\texttt{CheckMATE}\cite{Drees:2013wra,Barr:2003rg,Cheng:2008hk,Cacciari:2005hq,Cacciari:2008gp,Cacciari:2011ma,
deFavereau:2013fsa,Lester:1999tx,Read:2002hq}. The analysis implemented in the package have been validated by the code's authors.
We double checked by comparing the limits produced by the code with the official ones in three simplified models: 
chargino-neutralino production with $WZ$-mediated decay into leptons; 
chargino-neutralino production with slepton-mediated decay into leptons;
and left-handed slepton pair production. 
In all cases we found excellent agreement with the published results.

The CMS 3-lepton search\cite{Khachatryan:2014qwa} is recast using the code designed by some of us and described in detail in 
Refs.\cite{Fowlie:2012im,Fowlie:2013oua,Kowalska:2013ica}. For every point in the considered parameter space a set of $10^5$ events 
is generated at the parton level with \texttt{\pythia8}\cite{Sjostrand:2007gs}, and the hadronization products are passed 
to the fast detector simulator \texttt{\delphes}\cite{deFavereau:2013fsa} to reconstruct the physical objects. 
The CMS detector card is used, with the settings adjusted to those recommended by the experimental collaboration. 
Two kinematical variables proper of the 3-lepton search, invariant mass $M_{ll}$ and transverse mass $M_T$, are then constructed and used 
to divide the signal events into exclusive kinematical bins defined in the experimental paper. 
Finally, the acceptances/efficiencies are calculated as the fraction of all generated events that pass the applied cuts. 
The number of signal events is calculated as the product of the efficiency, luminosity and cross-section, where we use the NLO+NLL 
cross-sections provided by the LHC SUSY Cross Section Working Group\cite{LHCSXSECWG}. 

The exclusion bounds are set according to a marginalized likelihood ratio method, 
with the Poisson likelihood distribution marginalized over the background and normalized to the background-only hypothesis as described in 
Sec.~3 of\cite{Kowalska:2013ica}.
The total likelihood ${\cal L_{\textrm{LHC}}}$ is given by the product of the likelihoods from each signal bin.

Note, incidentally, that the ATLAS 2-lepton search uses instead signal regions that are not mutually exclusive.
In that case, only the likelihood from the signal bin with the largest expected sensitivity is taken. 
As was mentioned above, when we combine the limits of different 3-lepton searches we assume that a point 
is excluded at the 95\% \cl\ when $-2 \log{\cal L_{\textrm{LHC}}} > 5.99$ for at least one of the searches.

\begin{figure}[t]
\centering
\subfloat[]{%
\label{fig:a}%
\includegraphics[width=0.47\textwidth]{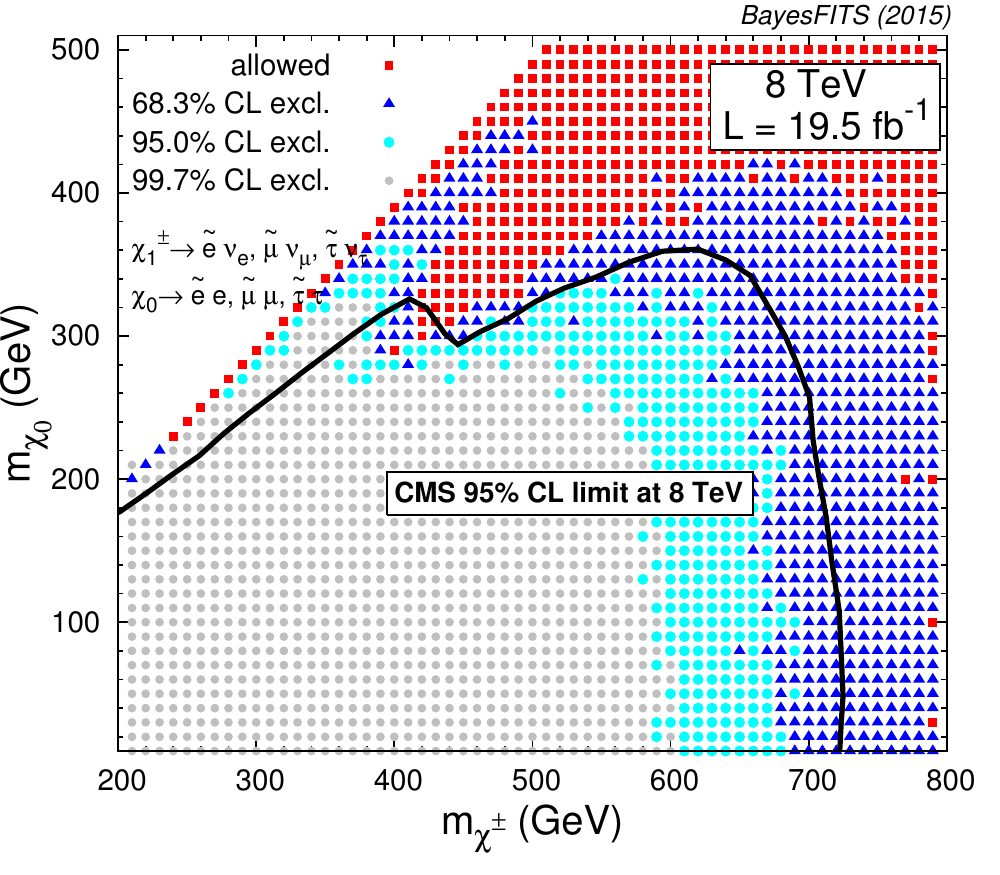}
}%
\hspace{0.02\textwidth}
\subfloat[]{%
\label{fig:b}%
\includegraphics[width=0.47\textwidth]{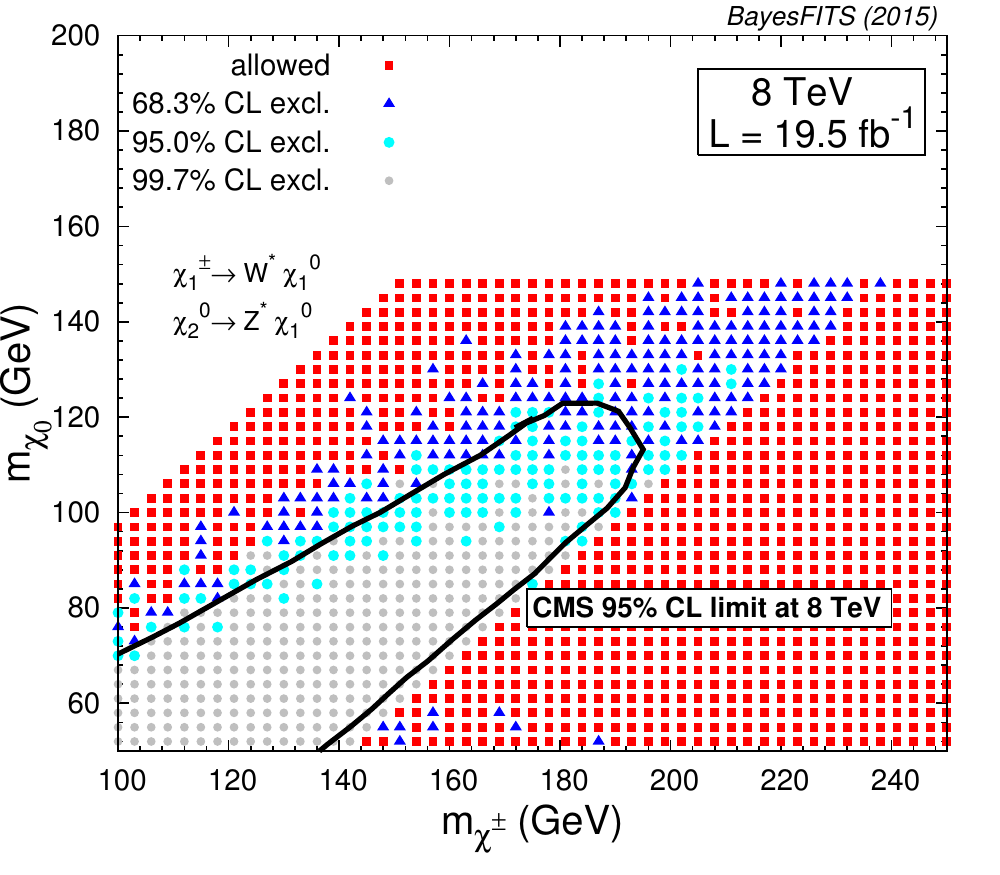}
}%
\caption{\footnotesize Exclusion confidence levels given by our 3-lepton likelihood function compared to the official 
CMS 95\%~C.L. exclusion bound\cite{Khachatryan:2014qwa} shown as a black solid line. 
\protect\subref{fig:a} SMS case of chargino-neutralino 
production with slepton-mediated decays in the ``flavor-democratic'' scenario. \protect\subref{fig:b} SMS case with with no intermediate sleptons decays. 
Gray dots are excluded at the 99.7\%~C.L., cyan circles at the 95.0\%~C.L., and blue triangles at the 68.3\%~C.L. Red squares indicate all other points.}  
\label{fig:CMS_SMS_validation}
\end{figure}

We present in \reffig{fig:CMS_SMS_validation} the validation of the limits obtained with our code by comparing them to the official CMS 
95\%~C.L. bounds. In \reffig{fig:CMS_SMS_validation}\subref{fig:a} we show the SMS case of chargino-neutralino 
production with slepton-mediated decays in the ``flavor-democratic'' scenario, 
while in \reffig{fig:CMS_SMS_validation}\subref{fig:b} we show the results for the case with no light sleptons. 
The points excluded by the likelihood function at the 99.7\%~C.L. are shown as gray dots, those excluded at the 95.0\%~C.L. as cyan circles, 
and those excluded at the 68.3\%~C.L. as blue triangles. Red squares indicate all other points. 
The solid black lines show the official CMS 95\%~C.L. exclusion limits from\cite{Khachatryan:2014qwa}. 

Besides the present LHC bounds we also calculate the sensitivity of the ATLAS 2-lepton and CMS 3-lepton searches at the LHC 14\tev\ run. 
We assume $L=300\invfb$ integrated luminosity. In each case we simulate the dominant SM backgrounds. 
For the 3-leptons search these are $WZ$ and $t\bar{t}$ production, as well as rare SM processes such as $t\bar{t}Z/W/H$ and triboson production. 
For the 2-lepton search the dominant backgrounds come from diboson production and $t\bar{t}$ production. 
Background events are generated at the LO using \madgr\cite{Alwall:2014hca} and showered using \texttt{\pythia8}. 
The cross-sections are calculated at the NLO using \madgr. 
We generated $1.5\times 10^6$ $t\bar{t}$ events, $2\times10^5$ $WZ$ events, and $3\times 10^5$ rare SM process events for the 3-lepton search. 
For the 2-lepton search we additionally simulated $10^6$ $W^+ W^-$ events and $10^6$ $Z Z$ events. 
The efficiencies for the background samples are derived applying the same experimental cuts used for the 8\tev\ run, 
and the number of background events is calculated as the product of the efficiency, luminosity and cross-section. 
The uncertainty in the number of background events is evaluated as the sum in quadrature of two terms: the uncertainty of the 
cross-section determination, given by \texttt{MadGraph}; and the statistical uncertainty of the efficiency 
determination with the Monte Carlo 
simulation.\footnote{The uncertainty of the efficiency $\epsilon$ is defined as $\sigma_{\epsilon}=\sqrt{\frac{(1-\epsilon)\epsilon}{N-1}}$, 
where $N$ is the total number of events generated in a Monte Carlo simulation. 
If the efficiency is equal zero, $\sigma_{\epsilon}$ is reduced to  $\sigma_{\epsilon}=1/N$.}
Note that this approach leads to a conservative estimate of future sensitivity, 
as one expects the eventual uncertainties determined by the experimental collaborations using data-driven methods to be several times smaller.

The background generation is first validated at 8\tev\ by comparing the Monte Carlo results to the number of expected background events 
given in the experimental papers. In the course of the validation procedure 
the exclusion bounds for the ``flavor democratic'' and left-handed slepton SMS 
were also rederived using our background determination. 
In both cases we obtained good agreement of our procedure with the official experimental results.

The projected exclusion bounds at 14\tev\ are obtained by setting the number of observed events equal to the expected number of background events. 
In the ATLAS 2-lepton search discrimination between the signal and background yields is performed by means of the 
kinematical variable $m_{T2}$\cite{Lester:1999tx,Barr:2003rg}, 
with largest values of $m_{T2}$ probing large mass splittings between the slepton and neutralino. 
In the 8\tev\ analysis the largest $m_{T2}$ considered is $m_{T2} > 150$\gev. However, 
when the mass of the slepton increases, the $m_{T2}$ distribution for the signal falls more slowly 
than for the background well beyond this cut. Since at 14\tev\ we expect to probe much larger slepton masses than in the 8\tev\ case,
we enhance the signal region by adding two new inclusive bins, $m_{T2} > 260$\gev\ and $m_{T2} > 310$\gev, to increases the sensitivity of 
the search in the high mass region.

\begin{figure}[t]
\centering
\subfloat[]{%
\label{fig:a}%
\includegraphics[width=0.47\textwidth]{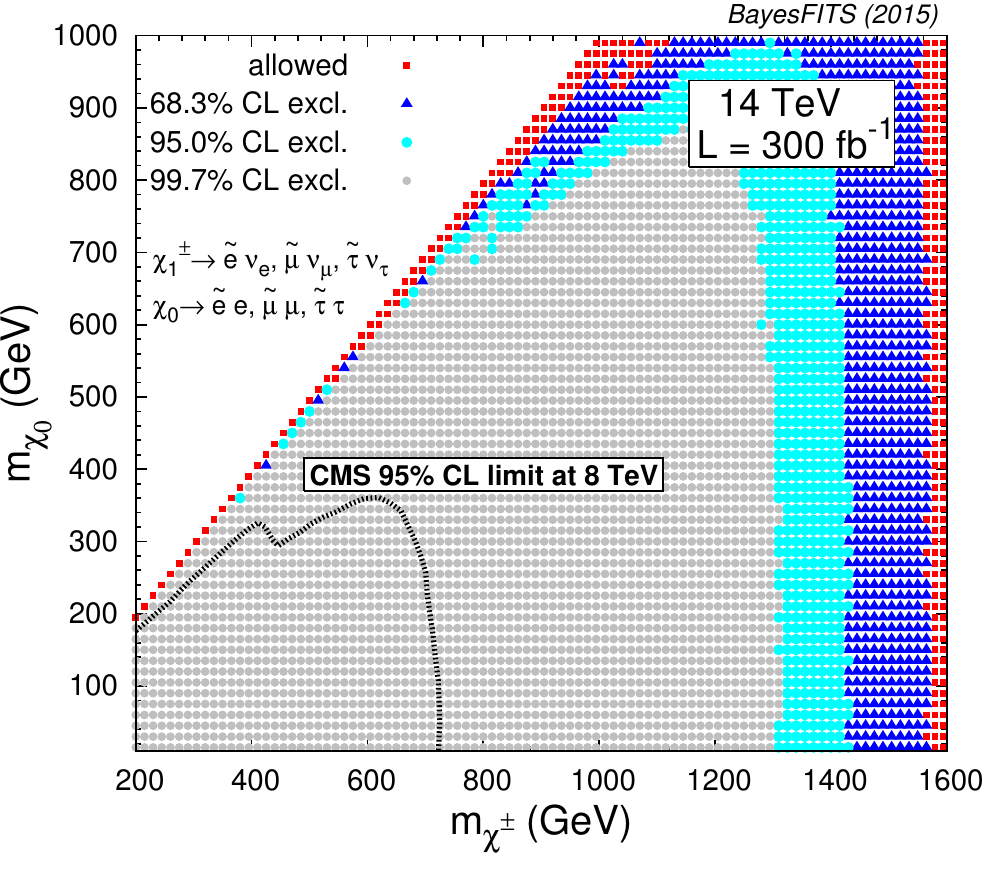}
}%
\hspace{0.02\textwidth}
\subfloat[]{%
\label{fig:b}%
\includegraphics[width=0.47\textwidth]{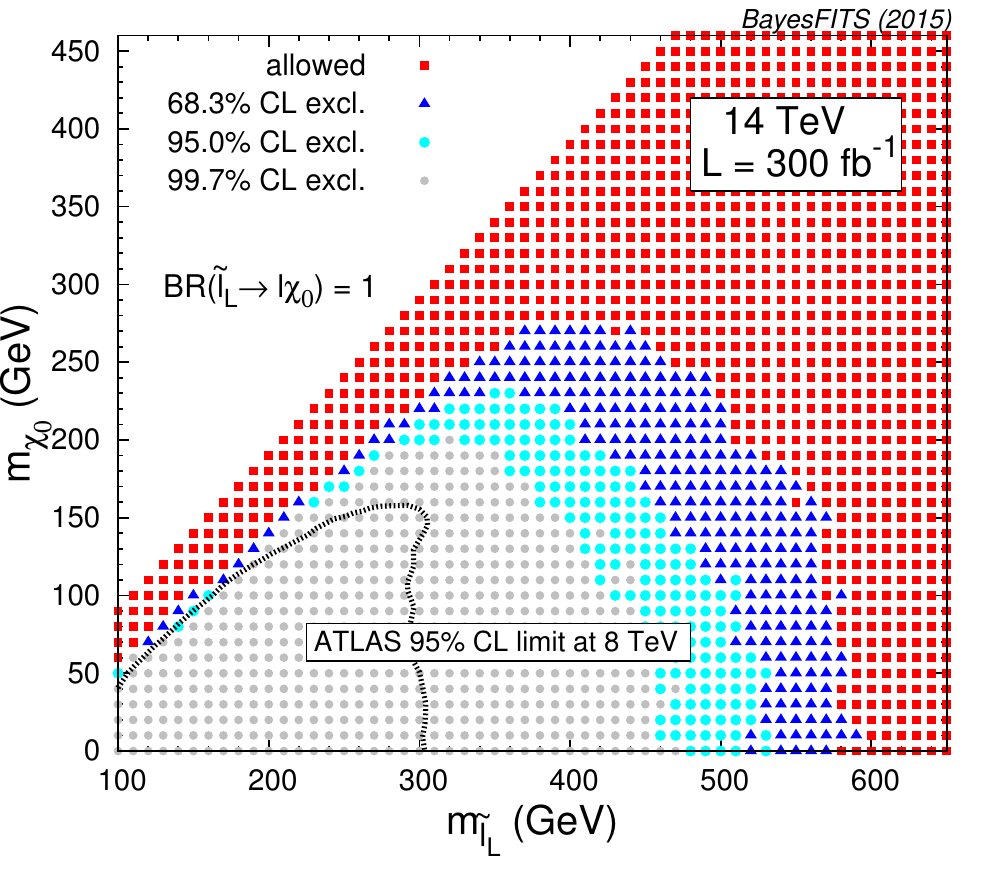}
}%
\caption{\footnotesize Projected exclusion bounds for the LHC 14\tev\ run with 300\invfb. 
\protect\subref{fig:a} 3-lepton search for $\charone \neuttwo$ production with intermediate sleptons in the ``flavor-democratic'' SMS at CMS.
\protect\subref{fig:b} 2-lepton search for $\tilde{e}_L \tilde{e}_L$ production at ATLAS. The color code is the same as in \reffig{fig:CMS_SMS_validation}.
The bounds from the 8\tev\ run are shown as dotted black lines.}  
\label{fig:LHC14_valid}
\end{figure}

In \reffig{fig:LHC14_valid}\subref{fig:a} we present the projected 14\tev\ sensitivity of the 3-leptons CMS search for chargino-neutralino 
production with slepton-mediated decays in the ``flavor-democratic'' scenario, with integrated luminosity $L=300\invfb$. 
The color code is the same as in \reffig{fig:CMS_SMS_validation}. 
We calculate that the lower bound on the chargino mass for a neutralino LSP lighter than $\sim 900\gev$ 
can be extended up to $\sim 1400\gev$, which is a factor of two increase with respect to the 8\tev\ result. 
In \reffig{fig:LHC14_valid}\subref{fig:b} we show the sensitivity of the ATLAS 2-lepton search for left-handed slepton pair production SMS. 
The difference in shape between the 8\tev\ and 14\tev\ limits is due to the extra bins at large $m_{T2}$, 
clearly indicating the importance of these bins in adding sensitivity in the heavy slepton region.

\subsection{Limits on GUT scenarios from the LHC 8~TeV run\label{sec:8tevlhc}}

We show here the present LHC bounds on the parameter space of the models of \refsec{sec:scenarios}. We obtain them by applying 
the simulation of the searches described in \refsec{sec:searches} to the model scans.
For each point we simulate $10^5$ events at LO using \texttt{\pythia8} for each of the relevant production mechanisms: 
$\neuttwo \charone$ and  $\tilde{\chi}_{1,2}^\pm \tilde{\chi}_{3,4}^0$ for the 3-lepton searches and 
additionally $\tilde{l}_L^+ \tilde{l}_L^-$, $\tilde{l}_R^+ \tilde{l}_R^-$, $\tilde{l}_L^{\pm} \tilde{\nu}_l$, $\tilde{\nu}_l \tilde{\nu}_l^*$ and $\tilde{\chi}_1^+ \tilde{\chi}_1^-$ for the 2-lepton search.
Cross sections are calculated at NLO using \madgr.

\begin{figure}[t]
\centering
\subfloat[]{%
\label{fig:a}%
\includegraphics[width=0.47\textwidth]{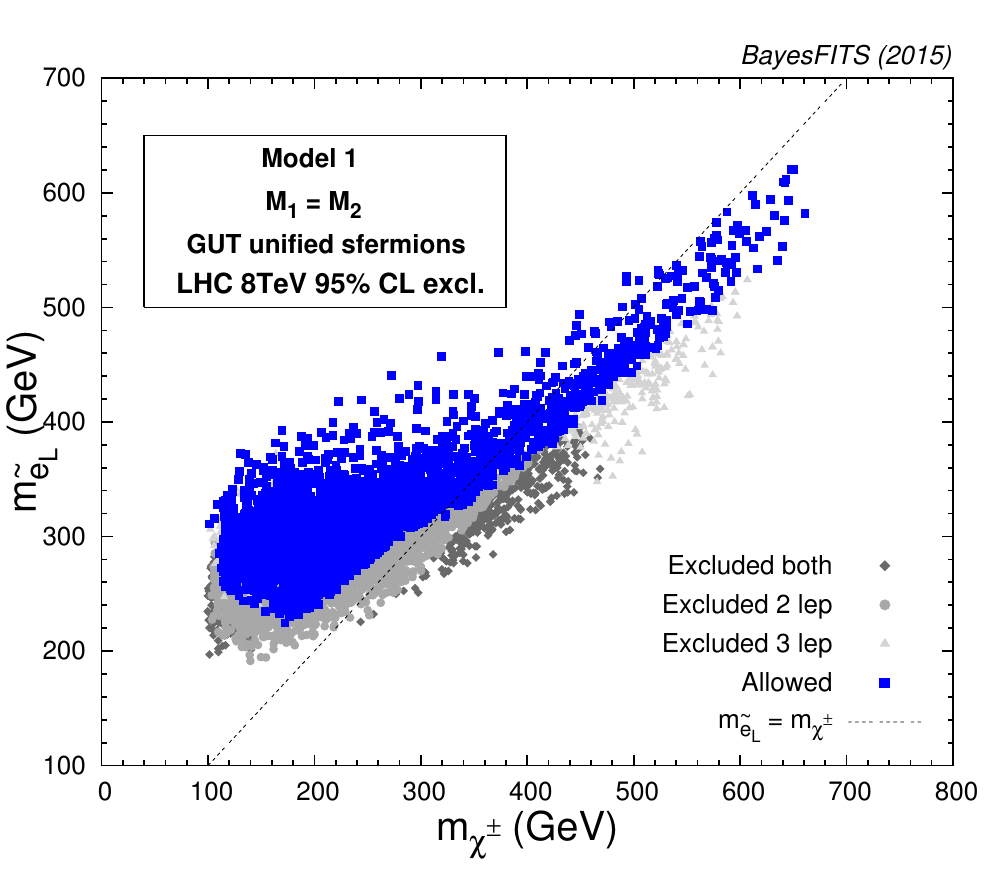}
}%
\hspace{0.02\textwidth}
\subfloat[]{%
\label{fig:b}%
\includegraphics[width=0.47\textwidth]{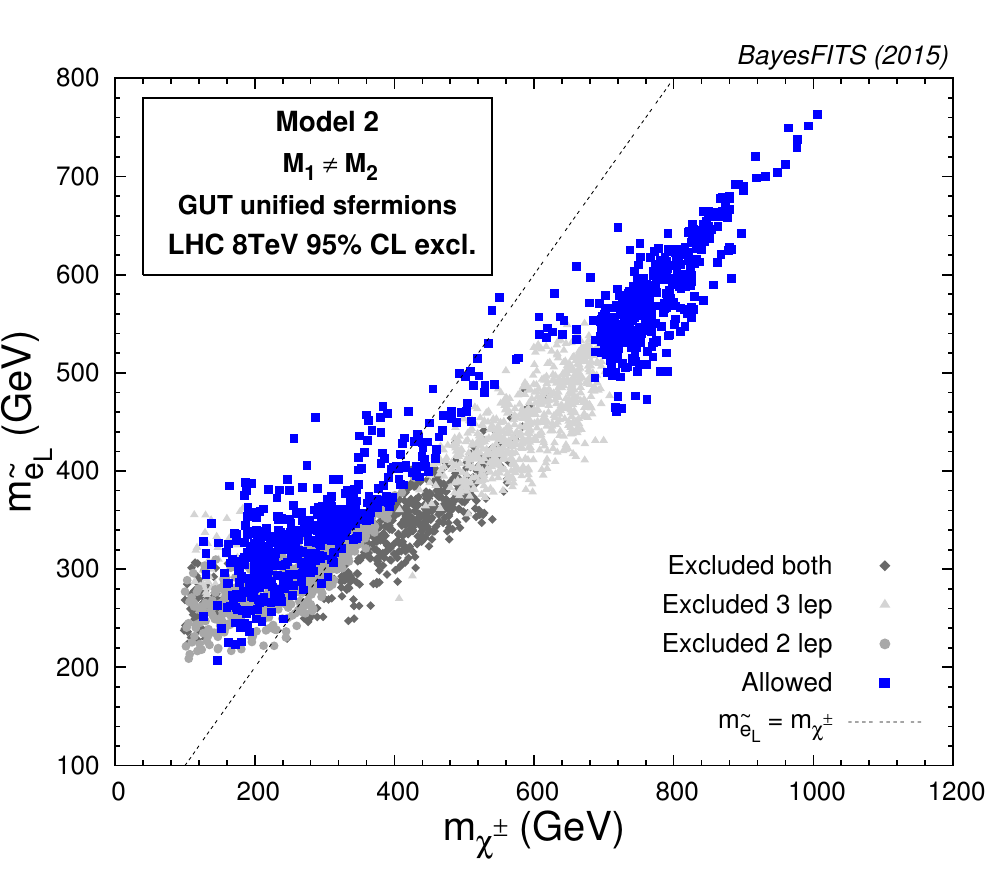}
}%

\subfloat[]{%
\label{fig:c}%
\includegraphics[width=0.47\textwidth]{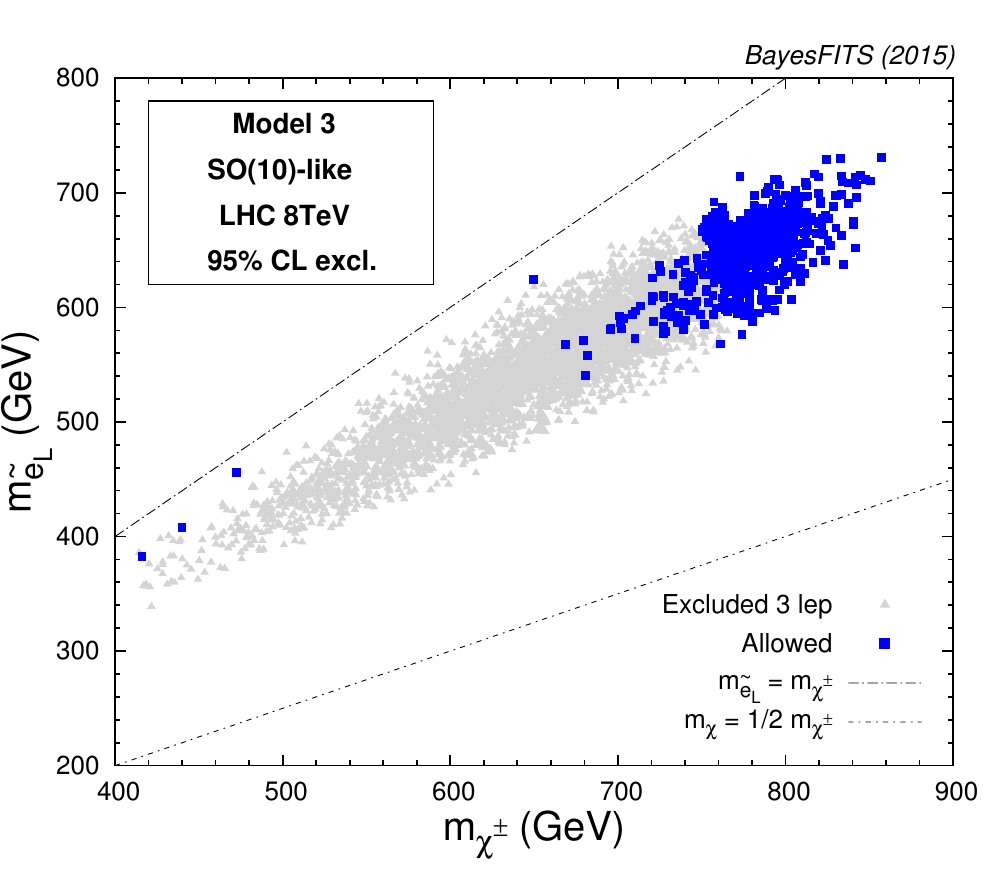}
}%
\hspace{0.02\textwidth}
\subfloat[]{%
\label{fig:d}%
\includegraphics[width=0.47\textwidth]{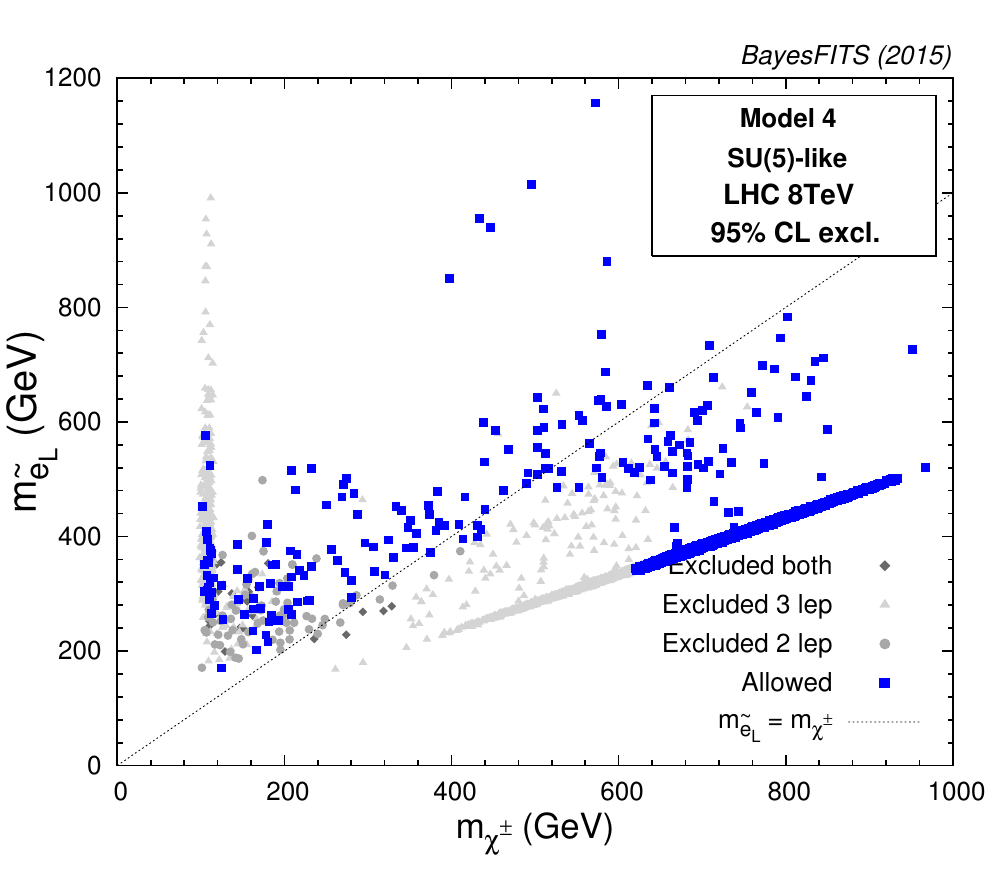}
}%
\caption{\footnotesize \protect\subref{fig:a} The bounds from the LHC 8\tev\ run on the points of Model~1 
in the ($m_{\charone}$, $m_{\tilde{e}_L}$) plane. Points excluded at the 95\%~C.L. by one of the implemented 3-lepton searches are shown as light gray triangles, 
points excluded by the 2-lepton search are shown as gray circles, and points excluded in both topologies are shown as dark gray diamonds. Blue squares represent the 
points still unconstrained at the LHC. 
\protect\subref{fig:b} Same as \protect\subref{fig:a} for Model~2. 
\protect\subref{fig:c} Same as \protect\subref{fig:a} for the part of the parameter space of Model~3 that is \textit{not} common to Model~1. \protect\subref{fig:d} 
Same as \protect\subref{fig:a} for Model~4.} 
\label{fig:8tev}
\end{figure}

We show in \reffig{fig:8tev}\subref{fig:a} the LHC bounds on the CMSSM-like scenario of Model~1. 
All points satisfy the constraints of \reftable{tab:exp_constraints} at the 95\%~C.L.
The points excluded at the 95\%~C.L. by the 3-lepton searches are shown as light gray triangles, those excluded by the 2-leptons searches as 
gray circles, and those excluded simultaneously in both topologies as dark gray diamonds. The points that are presently not excluded by the LHC 
are shown as blue squares. The plot is presented in the ($m_{\charone}$, $m_{\tilde{e}_L}$) plane, as the sensitivity 
of 3-lepton searches to the chargino mass increases in this plane from right to left and is thus 
orthogonal to the sensitivity of 2-lepton searches for slepton pair production, which increases instead from top to bottom.
Since $\mone=\mtwo=\mhalf$ at the GUT scale, the approximate relation
$\mchi\approx 0.5\,m_{\charone}$ holds for all shown points.
The exclusion criterion was described in \refsec{sec:searches}. The surviving points have 
$\Delchisq=\chisq-\chisq_{\textrm{min}}<5.99$, where $\chisq_{\textrm{min}}$ corresponds to the background only hypothesis.

One can identify different areas of exclusion from left to right in \reffig{fig:8tev}\subref{fig:a}. 
The few visible light gray triangles (and dark gray diamonds) on the left for $m_{\charone}\lesssim 120\gev$ are 
excluded by the 3-lepton search as \charone\ and \neuttwo\ give rise to 3-body decays: 
$e^{\pm}\chi e^{\mp}(\textrm{or }\nu_{e})$ or $\mu^{\pm}\chi \mu^{\mp}(\textrm{or }\nu_{\mu})$.
We found that some models with $m_{\charone} >120\gev$ predominantly show large decay rates to on-shell staus, 
$\neuttwo\charone\rightarrow \stau_1\tau\,\stau_1\nu_{\tau}$. 
As the chargino mass increases the search loses sensitivity due to the combined effect of the cross section drop
and the fact that the mass splitting between \neutone\ and $\stau_1$ is decreasing. 

The gray circles (and dark gray diamonds) for $m_{\tilde{e}_L}\lesssim 350\gev$ are excluded by the 2-lepton bounds 
on direct left- and right-slepton production. The 2-lepton search loses sensitivity when the slepton is much 
heavier than the chargino and the dominant decay channel then becomes $\slep\rightarrow \charone\nu_{l}$, 
which yields significantly softer final state leptons than those from $\slep\rightarrow l\chi$. 
On the other hand, an alternative channel opens up in this region: $\tilde{\nu}_l\rightarrow \charone\l^{\mp}$. 
Since the production cross section of a sneutrino is comparable to that of the corresponding slepton, 
many points are within the sensitivity of the 2-lepton search. As a result all the parameter space
with $m_{\tilde{e}_L}\approx m_{\tilde{\nu}_e}\lesssim250\gev$ is excluded.  

Finally, the region on the right of the plot, for $m_{\charone}>m_{\tilde{e}_L}$, 
is excluded by the 3-lepton searches, which are very efficient in detecting the 
$\neuttwo\charone\rightarrow \slep l\,\tilde{\nu}_l(\slep) l(\nu_l)$ topology,
unless the spectrum becomes excessively compressed in $m_{\charone}$ and $m_{\tilde{e}_L}$.

In \reffig{fig:8tev}\subref{fig:b} we show the equivalent exclusion plot for Model~2, which is characterized by 
$\mtwo\neq\mone$ at the GUT scale.
The LHC bounds in the ($m_{\charone}$, $m_{\tilde{e}_L}$) plane do not show great differences with Model~1, 
with the exception of the region of the parameter space corresponding to $m_{\charone}$ in the range $450-700\gev$, which is 
almost completely excluded by the 3-lepton searches. As a matter of fact, at the origin of the more elongated shape of Model~2's point 
distribution is the fact that the GUT-scale value of \mtwo\ is allowed to be larger than \mone, giving rise to heavier charginos in the stau-coannihilation region,
as heavy as $\sim 1\tev$ for some points. Selectron and chargino masses are, however, 
less compressed than in Model~1, so that 3-lepton searches are more sensitive to $m_{\charone}\lesssim700\gev$ for Model~2.

We showed in \refsec{sec:scenarios} that, when one considers GUT symmetry patterns that allow 
for disunifying the left and right-handed soft slepton masses, the allowed parameter space opens up and more
ways to obtain the correct relic density become viable.
Model~3 represents scenarios for which symmetry considerations allow for small deviations
from the universal case as happens, for example, in models of $SO(10)$ SUSY GUTs with a positive $D$-term mass.
In this case, besides the stau coannihilation and bulk regions, one finds right-handed slepton/neutralino coannihilation.
 
The LHC bounds on this additional part of the parameter space are shown in \reffig{fig:8tev}\subref{fig:c}. 
The exclusion limits are given in this region by the 3-lepton searches, as the 2-lepton search for direct 
slepton production is still not sensitive to $m_{\tilde{e}_L}\gsim 350\gev$.
On the other hand, this scenario provides optimal conditions 
for the 3-lepton search, with sleptons of both ``chiralities" such that $\mchi\lesssim m_{\slep_R} < m_{\slep_L}<m_{\charone}$,
so that the exclusion bounds are quite strong.
One can see that points with $m_{\charone}\lesssim 750\gev$ are excluded, with the exception of a few for which 
the bound is by $\sim100\gev$ weaker. Those are points characterized by 
$m_{\stau_1}< m_{\tilde{e}_L}$, so that the enhanced values of $\textrm{BR}(\charone\rightarrow\stau_1\nu_{\tau})$ 
and $\textrm{BR}(\neuttwo\rightarrow\stau_1\tau)$ have the effect of weakening the search's sensitivity.

We want to point out, however, that the GUT-scale boundary conditions are such that a light stau in the spectrum is
possible only for moderate values of \tanb, $\tanb\lesssim 25$. 
In our scan, \tanb\ is left a free parameter, but one ought to remember that
a successful $SO(10)$ unification pattern is likely to require large values of \tanb\
to achieve Yukawa coupling unification, even if one allows for the possibility of substantial threshold corrections (see, e.g., Ref.\cite{Anderson:1992ba}).
Thus, the points not excluded with $m_{\charone}\lesssim 750\gev$ are less theoretically motivated.

Finally, we show in \reffig{fig:8tev}\subref{fig:d} the LHC bounds on Model~4, which is
consistent with the $SU(5)$ pattern of unification and others that allow for similar low-scale behavior. 
The parameter space consistent with \gmtwo\ opens up significantly in this case and broader regions can evade the present bounds.
Besides the points where the mechanism for the relic density is right-handed slepton/neutralino coannihilation, 
as it was for Model~3, which are subject to the same bounds from the 3-lepton searches as the ones shown in \reffig{fig:8tev}\subref{fig:c}, 
one can also see a stripe of points where the left-handed sleptons are almost degenerate with the neutralino, 
for chargino masses in the range $200-950\gev$. 
Here the search limit is weakened to $m_{\charone}\lesssim 600\gev$. 
This is due to the fact that, while the left-handed sleptons are compressed with neutralinos, in the same region the right-handed sleptons 
also happen to be very close in mass to the charginos, thus strongly reducing the efficiencies. 

In the regions above the diagonal, for $m_{\charone}\gsim 200\gev$, the strength of the 3-lepton searches is weakened
by the fact that $m_{\tilde{e}_L} > m_{\charone}$. The dominant decay channels become $\neuttwo\charone\rightarrow h \neutone\, W^{\pm}\neutone$
for $m_{\charone}\simeq 450-600\gev$ and $\neuttwo\charone\rightarrow \stau_1\tau\,\stau_1\nu_{\tau}$ for $m_{\charone}\simeq 200-400\gev$.

Finally, for the solutions comprising the $h$-resonance region at $m_{\charone}\simeq100-120\gev$, $\mu$ is limited to values in the 
$300-800\gev$ range and stau-mixing is reduced so that for many points $m_{\tilde{\nu}_{\tau}}<m_{\charone}<m_{\stau_1}$.
The dominant decay channel, $\neuttwo\charone\rightarrow \tilde{\nu}_{\tau}\nu_{\tau}\,\tilde{\nu}_{\tau}\tau$, is invisible to 3-lepton searches. 
Pair production of sneutrinos of the first two generations do, however, decay to lepton + chargino and provide signatures within the reach of the 2-lepton search, but the sensitivity is not at the 
moment high enough to cover the whole region. One finds that it must be $m_{\tilde{\nu}_l}\approx m_{\slep_L}\gtrsim 240\gev$ for the points with $m_{\charone}\simeq 100-120\gev$.

\subsection{Projections for the LHC 14 TeV run}

In this section we investigate to what extent the second run at LHC will be able to explore the parameter space of the GUT-scale models defined in 
\refsec{sec:scenarios}. We do so by applying the likelihood function described at the end of \refsec{sec:searches} to the model points in all our scans.  
We remind the reader that we assume 14\tev\ center-of-mass energy and a target luminosity $L=300\invfb$.

\begin{figure}[t]
\centering
\subfloat[]{%
\label{fig:a}%
\includegraphics[width=0.47\textwidth]{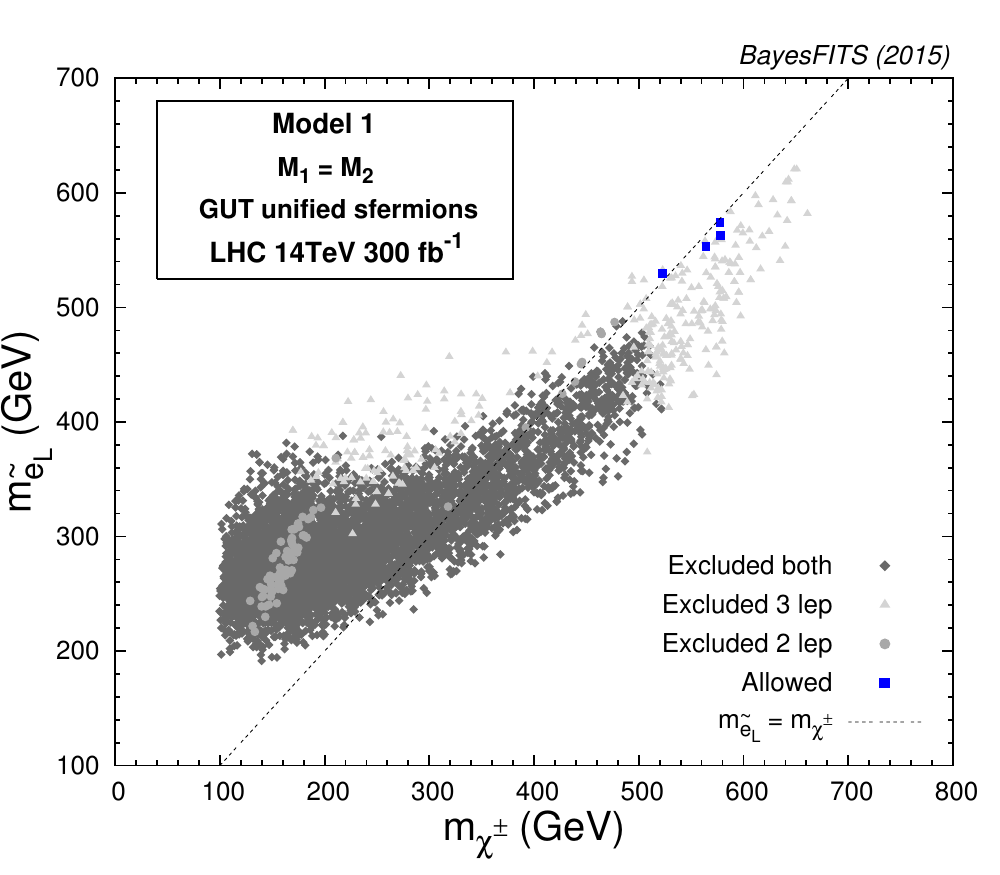}
}%
\hspace{0.02\textwidth}
\subfloat[]{%
\label{fig:b}%
\includegraphics[width=0.47\textwidth]{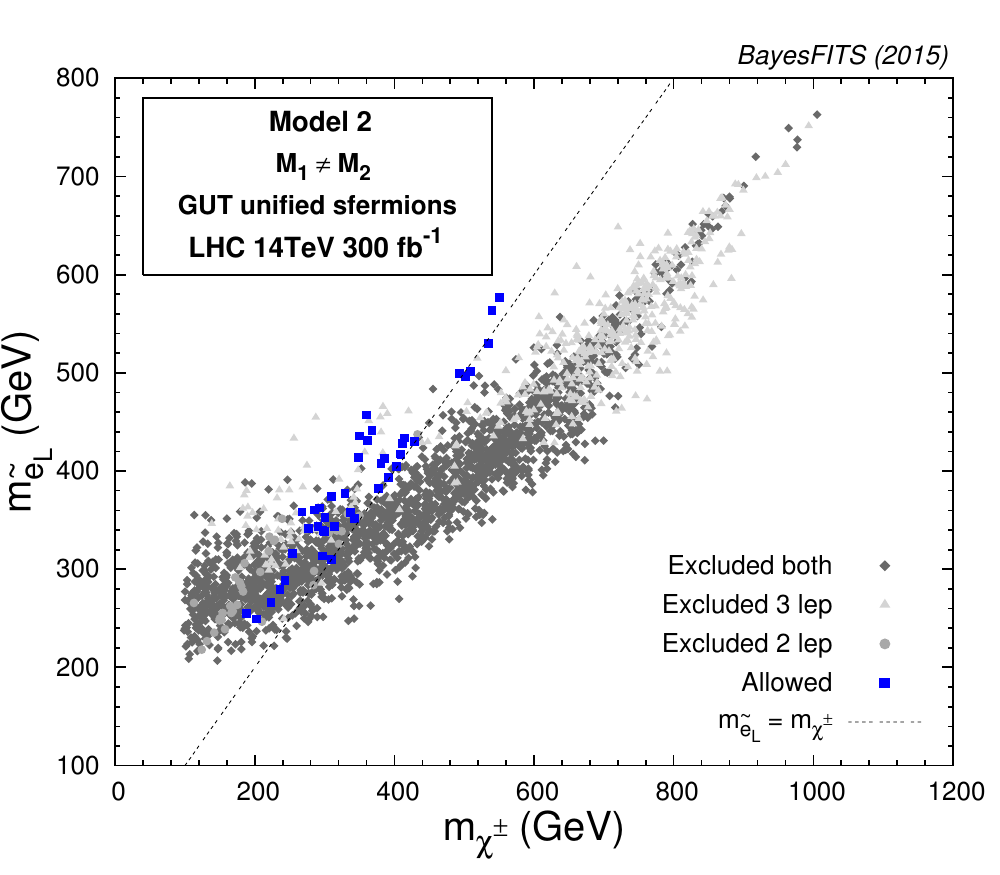}
}%

\subfloat[]{%
\label{fig:c}%
\includegraphics[width=0.47\textwidth]{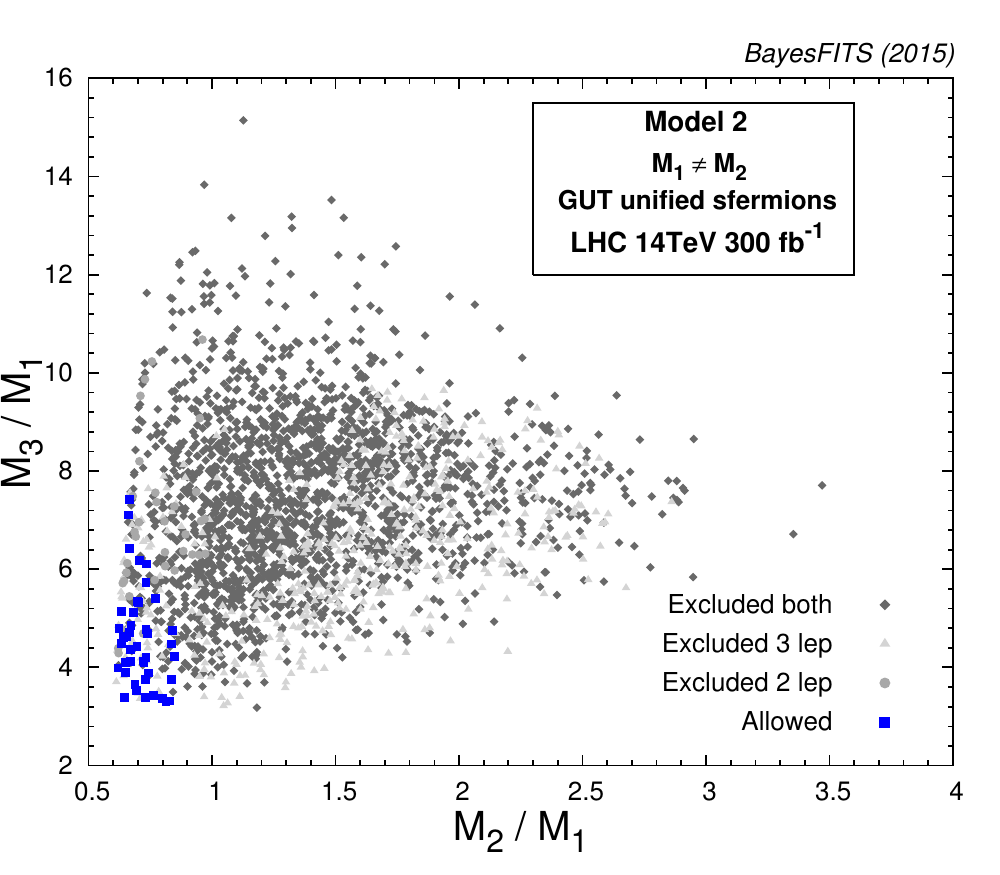}
}%
\hspace{0.02\textwidth}
\subfloat[]{%
\label{fig:d}%
\includegraphics[width=0.47\textwidth]{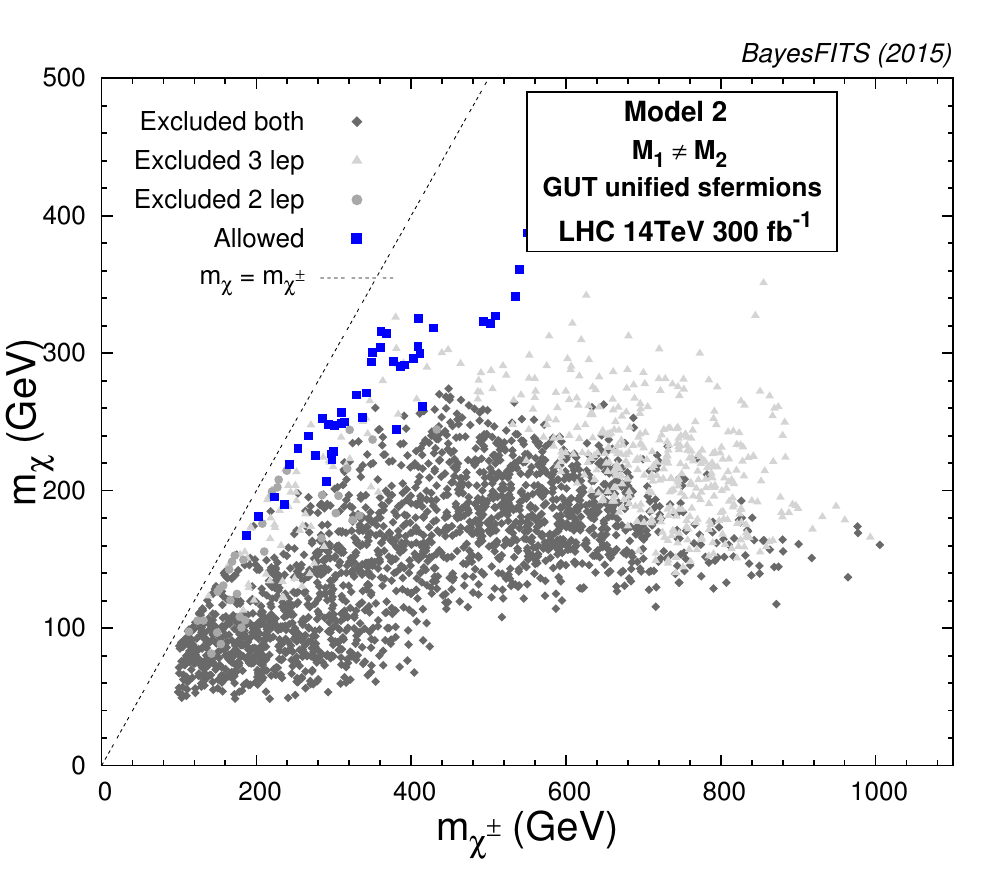}
}%
\caption{\footnotesize \protect\subref{fig:a} The expected reach of the LHC 14\tev\ run on the points of Model~1 
in the ($m_{\charone}$, $m_{\tilde{e}_L}$) plane. The color code is the same as in \reffig{fig:8tev}.
\protect\subref{fig:b} Same as \protect\subref{fig:a} for Model~2. 
\protect\subref{fig:c} The projections for Model~2 in the ($\mtwo/\mone$, $M_3/\mone$) plane. 
\protect\subref{fig:d} The projections for Model~2 in the ($m_{\charone}$, \mchi) plane.} 
\label{fig:14tev}
\end{figure}

In \reffig{fig:14tev}\subref{fig:a} we show the projected 95\% \cl\ bounds for Model 1. 
The color code is the same as in \reffig{fig:8tev}. 
Remarkably, the parameter space can be excluded in its near entirety, predominantly by the 3-lepton search.
It is interesting to note, however, that the regions of the parameter space for which the sensitivity of 3-lepton
searches will remain insufficient, will be covered in complementarity by 2-lepton searches.
This is the case, for example, of the points marked by gray circles at $m_{\charone}\simeq 100-180\gev$: 
their dominant decay chain is $\neuttwo\charone\rightarrow \stau_1\tau\,\stau_1\nu_{\tau}$, with  
stau and neutralino masses within $\sim20\gev$ from one another. The resulting tau is so soft 
that detection will probably be a challenge even at 14\tev. For the same points, however, 
the low-energy spectrum shows sleptons of the first two generations light enough to be easily detected in 
the 2-lepton search for direct pair production.
    
The only points in the plot that survive our simulation, shown as blue squares, lie in the compressed spectra region
$m_{\slep_L}\approx m_{\charone}\simeq 500-600\gev$. They also remain beyond the 95\% \cl\ reach  of the 2-lepton search, although our simulation 
places them within 68\% \cl\ reach. We remind the reader that our treatment of the 14\tev\ SM backgrounds is conservative. 
When a more precise background determination is provided by the experimental collaborations, this region may be tested entirely with an even lower luminosity.
 
If the gaugino mass universality condition is relaxed, as is the case of Model~2, new possibilities of evading the LHC exclusion bounds appear.
In \reffig{fig:14tev}\subref{fig:b} we show the projected bounds for Model~2.
The points outside of the LHC reach, indicated by blue squares, are characterized by a GUT-scale ratio $\mtwo/\mone<1$,
as can be seen in \reffig{fig:14tev}\subref{fig:c} where we show the LHC projection in the plane of the ratios
($\mtwo/\mone$, $M_3/\mone$).
These points feature chargino and neutralino masses quite close to each other, as shown
in \reffig{fig:14tev}\subref{fig:d} where we show the ($m_{\charone}$, \mchi) plane.
Note that the distribution in the ($m_{\tilde{e}_L}$, \mchi) plane (which we do not present here) show very similar behavior, 
namely the surviving points lie also on the compressed region for 2-lepton searches, making this combination of parameters 
very challenging even for the 14\tev\ run. 

Recent studies that have looked into possible strategies for a more comprehensive coverage of 
SUSY spectra compressed in the EW sector at the LHC (and possibly the ILC) can be found, e.g., in 
Refs.\cite{Deppisch:2014aga,Baer:2014kya,Dutta:2014jda,Han:2014aea,Barr:2015eva}.
The projected reach with 300\invfb\ from, e.g., Ref.\cite{Baer:2014kya} is $m_{\charone}\simeq 250\gev$ for a pure higgsino, 
which seems to fall short of probing these scenarios entirely. 

Incidentally, \reffig{fig:14tev}\subref{fig:c} also shows that it is difficult to generate 
points characterized at the same time by $\mtwo/\mone<1$ and very large values of $M_3$ at the GUT scale, as 
two-loop effects due to the gluino mass in the RGEs make the lightest stau tachyonic at the low scale. Hence the reduced density 
of points for $M_3/\mone\gsim 8$.

\begin{figure}[t]
\centering
\subfloat[]{%
\label{fig:a}%
\includegraphics[width=0.47\textwidth]{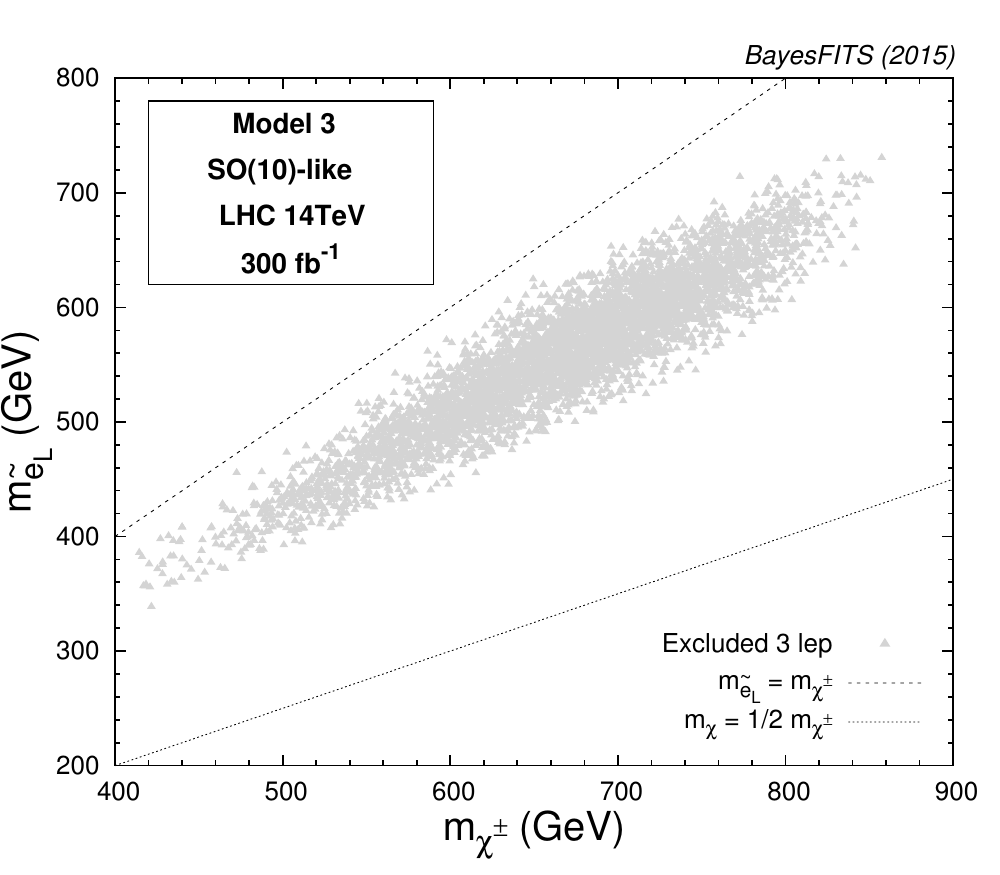}
}%
\hspace{0.02\textwidth}
\subfloat[]{%
\label{fig:b}%
\includegraphics[width=0.47\textwidth]{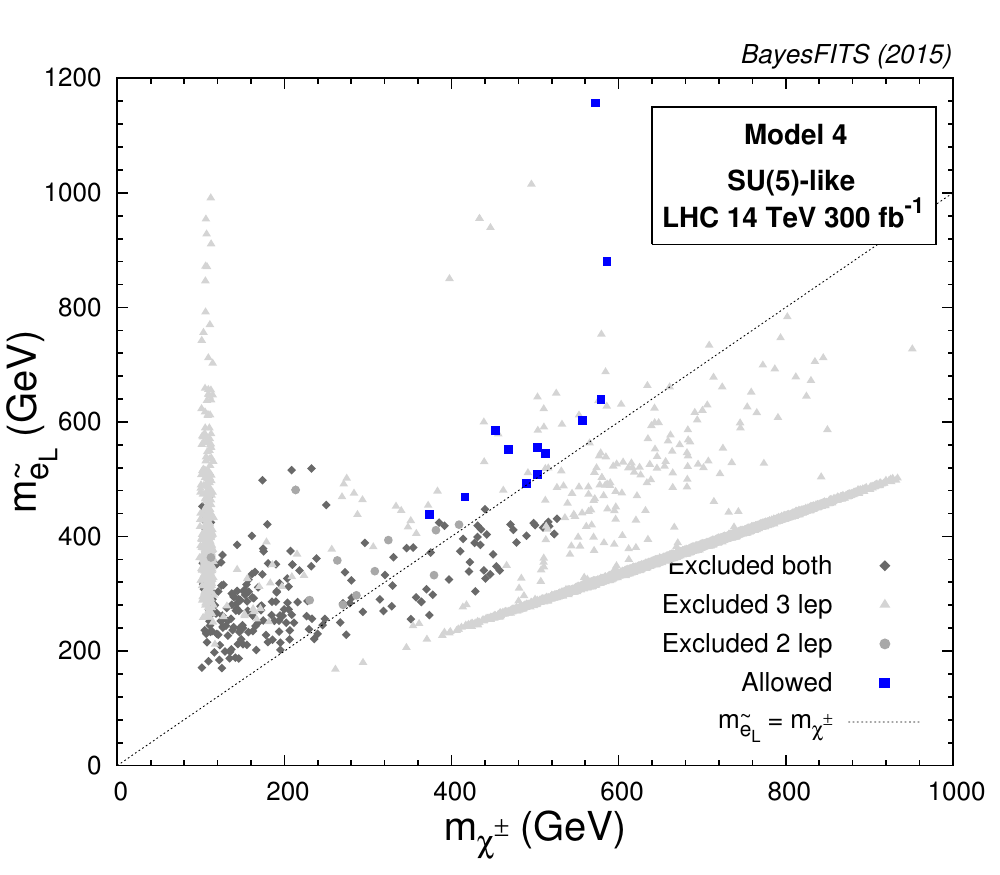}
}%
\caption{\footnotesize \protect\subref{fig:a} The expect reach of the LHC 14\tev\ run in the 
($m_{\charone}$, $m_{\tilde{e}_L}$) plane for the region of 
Model~3 that is \textit{not} common to Model~1.
\protect\subref{fig:b} The LHC 14\tev\ projections for Model~4. In both panels the color code is the same as in \reffig{fig:8tev}.} 
\label{fig:14sosu}
\end{figure}

Moving on to Model~3, the part of the parameter space due to right-handed slepton and neutralino coannihilation is going to be entirely probed
by 3-lepton searches with $\sim 100-110$\invfb\ of integrated luminosity. Figure~\ref{fig:14sosu}\subref{fig:a} shows that no point survives the cuts
in this region. 

More interesting is the case of Model~4.
We show in \reffig{fig:14sosu}\subref{fig:b} the projected 95\% \cl\ exclusion bounds for Model~4 in the ($m_{\charone}$, $m_{\tilde{e}_L}$) plane. 
The only part of the parameter space that remains unconstrained corresponds to points with $m_{\tilde{e}_L} > m_{\charone}$, 
characterized by large branching fractions for $\neuttwo \rightarrow \stau_1 \tau$ or $\neuttwo \rightarrow \neutone h$, 
depending on whether the lightest stau is lighter or heavier than \charone\ (and \neuttwo). 

The surviving points situated at $m_{\tilde{e}_L}\lesssim 600\gev$ should be in the future tested by the 2-lepton search even with $L=300\invfb$ 
in the likely event that the uncertainties in the background determination will be eventually smaller than our estimate. 
The points with $m_{\tilde{e}_L}> 600\gev$, obviously outside of the 2-lepton reach, are nonetheless characterized by large stau masses, 
and consequently larger branching fractions to the $\neuttwo\charone\rightarrow h \neutone\, W^{\pm}\neutone$ channel.  
They remain beyond the reach of the 3-lepton search at 300\invfb, 
but should eventually be tested with 3000\invfb\cite{ATL-PHYS-PUB-2014-010}.

To summarize our results, we show in \reftable{tab:benchmarks} the parameters and branching fractions of 3 benchmark points that survive our 14\tev\ projections.
Each point belongs to a different model. 

\begin{table}[t]
   \centering
   \begin{tabular}{|c|c|c|c|} 
      \hline
      Benchmark & BM1 & BM2& BM3 \\
      \hline
      \hline
      Model & \bf{Model 1} & \bf{Model 2} & \bf{Model 4} \\
       \hline
       \mzero & 406.7 & 280.0 & -- \\
       \hline
       $m_{10}$ & -- & -- & 269.8 \\
       \hline
       $m_5$ & -- & -- & 416.1 \\
       \hline
       \tanb\ & 22.9 & 8.0 & 18.4 \\
       \hline
       \mone\ & 747.6 & 556.4 & 759.4\\
       \hline
       \mtwo\ & $=\mone$ & 352.3 & $=\mone$\\
       \hline
       \mthree\ & 3427 & 2860 & 4227\\
       \hline
       \azero\ & 500 & $-9.3$ & 1572 \\
       \hline
       $\mhd^2$ & $=m_0^2$ & $=m_0^2$ & $-(5.07\times10^7)$\\
       \hline
       $\mhu^2$ & $=m_0^2$ & $=m_0^2$  &$-(5.01\times10^7)$ \\
       \hline
       \hline
      
         $m_{\neutone}$ & 304 & 219 & 306\\
	\hline	
       $m_{\charone}\approx m_{\neuttwo}$ & 577 & 243 &  579\\
       \hline
       $m_{\tilde{e}_L}$ & 574 & 288 & 639\\
       \hline
       $m_{\tilde{e}_R}$ & 498 & 354 & 310\\
	\hline
	$m_{\stau_1}$ & 329 & 233 & 772 \\
	\hline
	\hline
        				& $92\%$ $\stau\,\, \nu_\tau$		&  							& \\
        BF$(\charone)$	& $7\%$ $\tilde{\nu}_\tau\,\, \tau$	& $100\%$ $\stau\,\, \nu_\tau$		& $100\%$ $\neutone\,\, W$ \\
                                 	& $1\%$ $\tilde{\nu}_l\,\,l$			& 							& \\
       \hline
               			& $92\%$ $\stau\,\, \tau$			& 						& $93\%$ $\neutone\,\, h$\\
        BF$(\neuttwo)$ & $7\%$ $\tilde{\nu}_\tau\,\, \nu_\tau$& $100\%$ $\stau\,\, \tau$ 	& $6\%$ $\tilde{l}_R\,\,l$\\
                                 	& $1\%$ $\tilde{\nu}_l\,\,\nu_l$		& 						& $1\%$ $\neutone\,\, Z$\\
       \hline
                                    & 							& $54\%$ $\charone\,\, \nu_e$  	& $67\%$ $\neutone\,\, e$\\
        BF$(\tilde{e}_L)$ & 	$100\%$ $\neutone\,\, e$		& $28\%$ $\neuttwo\,\, e$		 	& $22\%$ $\neuttwo\,\, e$\\
                                    & 							& $18\%$ $\neutone\,\, e$	   		& $11\%$ $\charone\,\, \nu_e$\\
       \hline
       BF$(\tilde{e}_R)$& $100\%$ $\neutone\,\, e$		& $100\%$ $\neutone\,\, e$	& $67\%$ $\neutone\,\, e$\\
       \hline
                                      & 							& $51\%$ $\charone\,\, e$			& $70\%$ $\neutone\,\, \nu_e$\\
        BF$(\tilde{\nu}_e)$ & 	$100\%$ $\neutone\,\, \nu_e$	& $25\%$ $\neuttwo\,\, \nu_e$ 		& $20\%$ $\charone\,\, e$\\
                                      & 							& $24\%$ $\neutone\,\, \nu_e$	   	& $10\%$ $\neuttwo\,\, \nu_e$\\
       \hline
       \hline
       \chisq\ (3 lepton) & 1.8 & 0.1 & 0.29 \\
       \hline
       \chisq\  (2 lepton) & 2.3 & 2.0 & 2.56 \\
       \hline
   \end{tabular}
   \caption{\footnotesize The model parameters for the benchmark points. All dimensionful quantities are given in \gev\ or $\gev^2$. 
   Also shown are the masses and branching fractions of the relevant particles in the EW sector 
   as well as the $\chi^2$ values from the LHC searches at 14\tev.}
   \label{tab:benchmarks}
\end{table}

\section{Summary and conclusions\label{sec:conclusions}}

In this paper we have taken at face value the possibility that the $\sim 3\sigma$
anomaly in the measurement of \deltagmtwomu\ finds its origins in the MSSM.
If that is the case, the discrepancy should be to some extent confirmed by the 
New Muon g-2 experiment at Fermilab in the next few years.

Given the present limits on SUSY masses from direct measurements at the LHC 
and the measurement of the Higgs mass at $\mhl\simeq125\gev$, 
a SUSY spectrum in agreement with all experimental constraints should 
feature sparticles charged under color significantly heavier than
the ones only charged under the EW gauge groups, whose masses must be in general of the order of
a few 100s~GeV to be in agreement with the measured \deltagmtwomu.
This raises the question as to what extent spectra characterized by this kind of hierarchy are 
consistent with the limits from the first run at the LHC 
and how deeply they can be further probed by the oncoming 14\tev\ run. 

In generic parametrizations of the MSSM the issue has been examined in the literature and a brief reminder of the present 
LHC bounds on the parameter space subject to the additional requirement that the DM constraints are well satisfied by the lightest neutralino
is presented in the first part of this paper. It is shown
that large fractions of the parameter space can easily evade the present LHC limits and 
also lie beyond future sensitivity. 

On the other hand, realistic SUSY scenarios often present additional constraints on the parameter space due to 
the mechanism of SUSY breaking or additional symmetries.
In gravity-mediated, GUT-defined scenarios it has been known for a while that models with gaugino non-universality 
can be at the same time in agreement with \gmtwo, direct limits from the LHC on colored sparticles, 
and the Higgs mass measurement, if the GUT-scale value of the gluino soft mass $M_3$ is substantially larger than 
\mone\ and \mtwo.  
  
In this paper we exhaustively confronted the above scenarios 
with the exclusion bounds from direct SUSY searches at the LHC,
particularly when additional constraints that come from the relic density and $B$ physics observables are also taken into account.
We considered four types of GUT-scale models characterized by non-universal boundary conditions. 
In the first we assumed GUT-scale universality of all the sfermions soft masses, as well as of the gaugino mass parameters $M_1$ and $M_2$. 
The gluino mass parameter, on the other hand, was allowed to be much heavier to boost the squark masses up to the multi-\tev\ regime, 
as required by the measured value of the Higgs boson mass, while leaving the EW part of the spectrum relatively light 
to accommodate the measurement of \gmtwo. 
In this scenario the proper value of DM relic density is obtained through neutralino LSP coannihilation with the lightest stau. 

This feature persists even when the gaugino mass universality condition is relaxed at the GUT scale, 
$M_1\neq M_2$, which is the property of the second model we investigated. 
In the third model, slight relaxation of the universality condition on the sfermion masses, 
as happens within the framework of $SO(10)$ GUT scenarios, introduces the possibility of efficient coannihilation with the right-handed sleptons. 
Finally, when the left- and right-handed slepton soft masses at the GUT-scale are treated as independent parameters, 
as is the case for example in our fourth, $SU(5)$-inspired model, also left-handed slepton coannihilation and Higgs resonance annihilation mechanisms become available. 
 
To investigate the impact of the LHC searches on these models we have simulated 
two kinds of searches to explore the EW part of the spectrum 
and therefore look for experimental signatures with varying number of leptons in the final state. 
The existing 2- and 3-lepton searches at ATLAS and CMS were recast using the publicly available code 
\texttt{CheckMATE} and a similar tool developed by some of us.

On the other hand, the just started run II of the LHC, with a target center-of-mass energy of 14\tev, 
is expected to probe the EW sector of the MSSM much more efficiently. 
Thus, we derived predicted sensitivities of the 2-lepton and 3-lepton searches assuming a target luminosity 
$L=300\invfb$ and performing a detailed SM background simulation. 

Our analysis provides strong limits from the 8\tev\ run on the parameter space of GUT-constrained scenarios consistent with \gmtwo.
However, a large number of model points are shown to evade the limits, thus leaving ample room for the 
explanation of the \gmtwo\ anomaly within GUT-scale SUSY models. 

In this regard, we have shown in this work that the parameter space surviving the bounds from the 8\tev\ run 
falls within the sensitivity of the 14\tev\ run with 300\invfb\ projected luminosity virtually in its entirety. 
This opens up 
the interesting possibility that, if the \gmtwo\ anomaly is real and will be confirmed by future dedicated experiments,
explanation within a large class of well motivated SUSY models will give sure signatures at the LHC or, alternatively, these
models will have to be abandoned as an explanation for the \gmtwo\ anomaly. 

A few small regions proved to be difficult to test 
even at the end of the 14\tev\ run, and they should be given special attention.\medskip 

$\bullet$ All of these models present large fractions of the parameter space where the neutralino is bino-like and almost degenerate 
with the lightest stau. This channel is notoriously difficult to test, because of the soft nature of the produced taus.
Particular combinations of the input parameters have been shown to 
additionally conspire to push the chargino outside the reach of direct pair 
production and degenerate with the
left selectron, so that both 3-lepton and 2-lepton searches cannot be used as a handle.\medskip      

$\bullet$ In models with $\mone\neq\mtwo$ at the GUT scale not many combinations of \mone\ and \mtwo\  
allow one to obtain the correct relic density, and those that do 
feature in general $\mtwo/\mone\gsim 1$, 
so that they fall within the reach of 3-lepton searches.
However, some models can have $\mtwo/\mone< 1$, driving the EW spectrum to be highly compressed
even for a predominantly bino-like neutralino.\medskip  

$\bullet$ Models with boundary conditions consistent with $SU(5)$ or Pati-Salam
are characterized by larger freedom in the parameter space, making them closer 
to what happens in generic parametrizations of the MSSM with bino-like DM. 
The relic abundance constraint often requires at least one among the left- or right-handed 
selectron/smuon to be light and degenerate with the neutralino. 
However, the situation can arise where 
the other one and the lightest stau are several 100s GeV heavier than the slepton coannihilating.
In this case 2-lepton searches will be ineffective, as the model presents a compressed spectrum, and at the same time 
the sensitivity reach of 3-lepton searches will be curbed by the absence of intermediate, fairly light sleptons. 
$\neuttwo$ thus decays predominantly into the Higgs bosons, requiring a much larger 
integrated luminosity.\medskip

Increasing luminosity, reducing the background uncertainties and combining multiple searches statistically 
can be the first step to entirely probe the remaining model points. 
On the other hand, it is not excluded that SUSY hides exactly in ``pockets'' of the parameter space that are particularly challenging experimentally. 
Therefore, the effort should also be put in developing new search strategies 
that would have power to test such elusive spectra. 
In particular, we believe that finding efficient ways to increase the LHC sensitivity to compressed
tau final states is of utmost importance as these taus seem to be the curtain behind which some SUSY scenarios might be hiding.
\bigskip

\noindent \textbf{Acknowledgments}
\medskip

\noindent  We would like to thank Daniel Schmeier for email support on using \texttt{\checkm}.
  This work has been funded in part by the Welcome Programme
  of the Foundation for Polish Science. K.K. is supported by the EU and MSHE Grant
  No. POIG.02.03.00-00-013/09.
  L.R. is also supported in part by a Lancaster-Manchester-Sheffield Consortium for Fundamental Physics under STFC grant ST/L000520/1.
 The use of the CIS computer cluster at the National Centre for Nuclear Research is gratefully acknowledged. 
\bigskip

\bibliographystyle{JHEP}

\bibliography{g2MSSM}

\end{document}